\documentclass[conference]{IEEEtran}

\ifCLASSOPTIONcompsoc
  \usepackage[nocompress]{cite}
\else
  \usepackage{cite}
\fi

\usepackage{algorithm}
\usepackage{algorithmic}
\usepackage{amsfonts}
\usepackage{amsmath}
\usepackage{amssymb}
\usepackage{amsthm}
\usepackage{array}
\usepackage{booktabs}
\usepackage{colortbl}
\usepackage{dblfloatfix}
\usepackage{enumitem}
\usepackage{fontawesome5}  
\usepackage{graphicx}
\usepackage{longtable}
\usepackage{makecell}
\usepackage{mathtools}
\usepackage{mdframed}
\usepackage{multirow}
\usepackage{multicol}
\usepackage{pifont}
\usepackage{subcaption}
\usepackage{tabularray}
\usepackage{tabularx}
\usepackage[most]{tcolorbox}
\usepackage{tikz}
\usepackage{xspace}
\usetikzlibrary{positioning,arrows.meta,shadows}
\usepackage[nointegrals]{wasysym}
\usepackage[table,dvipsnames]{xcolor}
\usepackage[hidelinks]{hyperref}
\definecolor{RoyalBlue}{rgb}{0,0.2,0.6} 

\definecolor{cellhighlightblue}{RGB}{200, 230, 255} 
\definecolor{cellhighlightgray}{gray}{0.85}
\definecolor{ctrlgreen}{RGB}{223,240,216}
\definecolor{fixedred}{RGB}{250,228,228}

\definecolor{takeawayblue}{RGB}{43,72,114}   
\definecolor{rqorange}{RGB}{176,103,28}      
\definecolor{findinggreen}{RGB}{48,108,88}    
\definecolor{boxgray}{RGB}{247,247,245}      

\usepackage{circledsteps}

\newcommand{\hhline}{%
    \noalign {\ifnum 0=`}\fi \hrule height 1pt
    \futurelet \reserved@a \@xhline
}

\newcommand{\full}{\CIRCLE}       
\newcommand{\half}{\LEFTcircle} 
\newcommand{\none}{\Circle}       

\hypersetup{
    colorlinks=true,
    linkcolor={RoyalBlue},
    citecolor={Maroon}
}

\theoremstyle{plain}

\theoremstyle{definition}

\theoremstyle{remark}

\definecolor{ProgA}{HTML}{D9EAD3}  
\definecolor{ProgB}{HTML}{FFF2CC}  
\definecolor{ProgC}{HTML}{FCE5CD}  
\definecolor{ProgD}{HTML}{FCE0E0}  

\newcommand{\numchip}[2]{%
  \begingroup
  \setlength{\fboxsep}{1.1pt}%
  \colorbox{#1}{\strut #2}%
  \endgroup
}

\newcommand{\asra}[1]{\numchip{ProgA}{#1}} 
\newcommand{\asrb}[1]{\numchip{ProgB}{#1}} 
\newcommand{\asrc}[1]{\numchip{ProgC}{#1}} 
\newcommand{\asrd}[1]{\numchip{ProgD}{#1}} 

\newcommand{\scorea}[1]{\numchip{ProgA}{#1}} 
\newcommand{\scoreb}[1]{\numchip{ProgB}{#1}} 
\newcommand{\scorec}[1]{\numchip{ProgC}{#1}} 
\newcommand{\scored}[1]{\numchip{ProgD}{#1}} 

\newcommand{\highlightcell}[1]{%
    \cellcolor{cellhighlightgray}#1%
}

\usepackage[most]{tcolorbox}

\newcounter{takeaway}
\newcounter{rq}
\newcounter{finding}

\tcbset{
  sokbox/.style={
    enhanced,
    breakable,
    boxrule=0pt,
    frame hidden,
    sharp corners,
    colback=boxgray,
    left=6pt,
    right=6pt,
    top=5pt,
    bottom=5pt,
    before skip=6pt,
    after skip=6pt,
    before upper={\noindent},
  }
}

\newcommand{\sokboxlabel}[1]{%
  {\normalfont\bfseries #1 }%
}

\newenvironment{takeawaybox}[1][]{
  \refstepcounter{takeaway}
  \begin{tcolorbox}[
    sokbox,
    borderline west={1.5pt}{0pt}{takeawayblue},
    #1
  ]
  \sokboxlabel{Takeaway\,\thetakeaway:}%
}{
  \end{tcolorbox}
}

\newenvironment{rqbox}[1][]{
  \refstepcounter{rq}
  \begin{tcolorbox}[
    sokbox,
    borderline west={1.5pt}{0pt}{rqorange},
    #1
  ]
  \sokboxlabel{RQ\,\therq:}%
}{
  \end{tcolorbox}
}

\newenvironment{findingbox}[1][]{
  \refstepcounter{finding}
  \begin{tcolorbox}[
    sokbox,
    borderline west={1.5pt}{0pt}{findinggreen},
    #1
  ]
  \sokboxlabel{Finding\,\thefinding:}%
}{
  \end{tcolorbox}
}

\newcommand{\DeepSeek}{\textit{DeepSeek-V3}\xspace}
\newcommand{\GPTFourO}{\textit{GPT-4o}\xspace}

\newcommand{\LlamaTHREETHREE}{\textit{Llama-3.3-70B}\xspace} 
\newcommand{\GeminiThree}{\textit{Gemini-3.1-Flash}\xspace}

\newcommand{\ChatGLM}{\textit{ChatGLM}\xspace}
\newcommand{\LlamaTHREE}{\textit{Llama-3}\xspace}
\newcommand{\MISTRAL}{\textit{Mistral}\xspace}
\newcommand{\QWENTwo}{\textit{Qwen2}\xspace}
\newcommand{\VicunaSeven}{\textit{Vicuna-7b}\xspace}
\newcommand{\VicunaThirteen}{\textit{Vicuna-13b}\xspace}

\newcommand{\GemmaNine}{\textit{Gemma-2-9B}\xspace}

\begin{document}

\title{SoK: Intent-Oriented Systematization of Multi-Turn LLM Jailbreaks}

\author{
  \IEEEauthorblockN{
    Siyuan Li\IEEEauthorrefmark{1}, 
    Aodu Wulianghai\IEEEauthorrefmark{1}, 
    Zehao Liu\IEEEauthorrefmark{1}, 
    Xi Lin\IEEEauthorrefmark{1}, 
    Qinghua Mao\IEEEauthorrefmark{1},  
    Haoyu Li\IEEEauthorrefmark{2}, 
    Xiang Chen\IEEEauthorrefmark{3}, \\ 
    Siyuan Liang\IEEEauthorrefmark{4}, 
    Jun Wu\IEEEauthorrefmark{1}, 
    Jianhua Li\IEEEauthorrefmark{1}, 
    Dacheng Tao\IEEEauthorrefmark{4}
  }
  \IEEEauthorblockA{
    \IEEEauthorrefmark{1}Shanghai Jiao Tong University, 
    \IEEEauthorrefmark{2}University of Illinois Urbana-Champaign, \\
    \IEEEauthorrefmark{3}Zhejiang University, 
    \IEEEauthorrefmark{4}Nanyang Technological University \\
    \{siyuanli, melusine.wlhad, liuzehao, linxi234, mmmm2018, junwuhn, lijh888\}@sjtu.edu.cn, 
    haoyuli9@illinois.edu, \\ wasdnsxchen@gmail.com, \{siyuan.liang, dacheng.tao\}@ntu.edu.sg
  }
}

\IEEEoverridecommandlockouts
\makeatletter\def\@IEEEpubidpullup{6.5\baselineskip}\makeatother
\IEEEpubid{\parbox{\columnwidth}{
		Network and Distributed System Security (NDSS) Symposium 2027\\
		22--26 March 2027, Seoul, Republic of Korea\\
		ISBN 978-1-970672-09-1\\  
		https://dx.doi.org/10.14722/ndss.2027.240458\\
		www.ndss-symposium.org
}
\hspace{\columnsep}\makebox[\columnwidth]{}}

\maketitle

\begin{abstract}
Large Language Models (LLMs) are increasingly deployed in interactive settings, where user intent commonly unfolds through multi-turn dialogue.
Multi-turn jailbreaks exploit this pattern by advancing a harmful intent across turns, so that no single message exposes the full objective.
However, existing work treats these attacks as a loose collection of prompt patterns and does not analyze how the adversary organizes and advances harmful intent across an interaction.
We develop an intent-oriented taxonomy that organizes multi-turn jailbreaks along four intent-organization dimensions.
Through controlled ablations, we find evidence that effectiveness depends more on how deliberately intent is organized across turns than on matched increases in context length or query count.
We further show that the way intent is organized determines the level at which it becomes detectable, pushing the required observation scope outward from the turn level to the session level to the cross-session level.
These findings indicate that turn-local safety mechanisms are structurally insufficient and that single-point evaluation overlooks how intent is organized, motivating evaluation protocols aligned to the level at which harmful intent becomes observable.
The code is available at: \url{https://github.com/SiyuanLi00/INTACT}.
\end{abstract}

\IEEEpeerreviewmaketitle

\section{Introduction}
Large Language Models (LLMs) have rapidly advanced in reasoning, multimodal understanding, long-context capabilities, and tool use. Recent frontier systems continue to expand the capability boundary of general-purpose AI, while LLMs are increasingly deployed in interactive applications, including conversational assistants~\cite{ouyang2022training, wu2024longmemeval, guan2026evaluating}, coding copilots~\cite{pearce2022asleep, sandoval2023lost, jiang2026survey}, enterprise workflows~\cite{greshake2023not, fan2024workflowllm}, and agentic systems~\cite{zhang2025agent, wu2024isolategpt}. 
This deployment shift changes the adversarial surface facing LLM-based systems: attacks are no longer confined to isolated benchmark inputs, but can unfold through dialogue history, tool-mediated workflows, and stateful interactions~\cite{debenedetti2024agentdojo, shahriar2025survey, deng2025ai, wei2023jailbroken, wang2025manipulating}. 
As LLMs move from controlled evaluation settings into real-world use, their growing capability and reach also create new avenues for misuse and adversarial manipulation~\cite{lin2024malla, shen2025gptracker, yu2024don}.

One of the most pressing forms of such misuse is eliciting harmful or policy-violating outputs from LLMs through adversarial user inputs~\cite{yao2024llm_security_privacy, roy2024chatbots}. To mitigate this risk, modern LLMs are typically subjected to safety alignment procedures~\cite{lu2025adversarial,xiao2025detoxifying} spanning training-time preference optimization~\cite{ouyang2022training, bai2022constitutional, zhao2025improving, qi2024fine}, adversarial red-teaming and safety data augmentation~\cite{mazeika2024harmbench, jiang2024wildteaming}, and inference-time defensive mechanisms~\cite{zou2024improving, zhang2024parden, pang2026safesteer}. Yet alignment remains imperfect under adaptive adversaries~\cite{zhang2024large, he2024you, yu2024llm, yu2025mind, gong2025safety, song2025refusal, su2024mission}. Jailbreak attacks exploit these imperfections by crafting prompts or interaction sequences that induce harmful or policy-violating behavior~\cite{yi2024jailbreaksurvey, yu2024don, xu2024comprehensive}. Prior attacks have evolved from manually crafted adversarial prompts~\cite{shen2024anything, krauss2025twinbreak}, to automated optimization and search techniques~\cite{zou2023universal, chao2025jailbreaking, mehrotra2024tree, liu2024making, li2024semantic}, and further to cross-modal exploits against vision-language models~\cite{yang2024sneakyprompt, dong2025fuzz, liu2025modifier, mao2025llms, villa2025exposing, hakim2026jailbreaking, ba2024surrogateprompt, ying2024jailbreak}. However, much of this line of work still treats the prompt, image, or individual instance as the primary unit of attack~\cite{chao2024jailbreakbench, shen2024anything, mazeika2024harmbench}, leaving the interaction process itself comparatively under-analyzed.

Multi-turn jailbreaks expose this missing interaction-level attack surface. 
In practical LLM use, user intent often unfolds over multiple turns, and adversaries can exploit the same interface by advancing a harmful objective gradually, adapting to refusals, and conditioning later requests on earlier model responses~\cite{laban2025llms, zhang2025survey, li2024llm, zou2024improving, russinovich2025great}. 
Existing methods instantiate this idea through escalation-based dialogues~\cite{russinovich2025great, weng2025foot, Cheng2024, Sun2024, ying2025reasoning}, adaptive attacker agents~\cite{deng2023masterkey, pavlova2024automated, chen2025strategize}, and multi-path or cross-session attack designs~\cite{zhou2025tempest, rahman2025x, Srivastav2025Safe, Wahreus2025Prompt}.
Despite their surface diversity, these attacks share a common structure: the harmful objective is not necessarily exposed in a single prompt, but is organized across turns, trajectories, or sessions. 
This makes a multi-turn jailbreak qualitatively different from simply repeating or rephrasing a single malicious request, and motivates treating it as an interaction-level threat model in its own right.

The rapid growth of jailbreak methods has prompted extensive efforts to systematize the surrounding literature. 
Some surveys and SoK studies examine LLM safety and security at a broad system level~\cite{shayegani2023llm_vulnerabilities_survey, yao2024llm_security_privacy, wang2025sok}, while others focus on prompt-level threats and defenses~\cite{knowlton2026prompt}. 
A growing body of jailbreak-specific work further studies attack taxonomies, guardrail design, and robustness evaluation~\cite{yi2024jailbreaksurvey, xu2024comprehensive, mao2025llms, hong2025sok, hakim2026jailbreaking, li2026honeytrap, xu2026sok}. 
Together, these works cover a spectrum from general LLM attack and defense surfaces to specialized jailbreak evaluation protocols. 
However, when multi-turn attacks are discussed, they are still typically organized by surface techniques, attacker capabilities, or evaluation outcomes, rather than by how adversaries structure harmful intent across an interaction. 
This leaves two central questions underexplored: how adversaries organize and advance harmful intent across turns, trajectories, and sessions; and at what observation scope such intent becomes detectable to a defender. 
These questions call for a systematization organized by the control point through which harmful intent is planned, distributed, and made observable.

To address these questions, we provide a comprehensive taxonomy of coordinated multi-turn jailbreaks in this SoK. 
Rather than treating these attacks as a disjointed collection of surface-level prompt templates, we identify four recurring dimensions along which adversaries organize harmful intent while planning and executing long-horizon attacks.
Based on these adversarial-strategy variations, we propose a unified taxonomy organized around \textbf{Dialogue-escalation Attacks} (\textbf{DEA}), \textbf{Strategy-refinement Attacks} (\textbf{SRA}), \textbf{Trajectory-branching Attacks} (\textbf{TBA}), and \textbf{Session-splitting Attacks} (\textbf{SSA}).
The dimensions are non-exclusive at the method level, so a method may receive multiple labels when its attack process substantively instantiates multiple control points.

Building on this taxonomy, we conduct targeted empirical analyses on representative DEA and SRA methods, where available implementations allow controlled ablations of single-trajectory attack mechanisms. 
These experiments test whether multi-turn jailbreak effectiveness is driven by accumulated context and query count, or by the deliberate organization of harmful intent across the interaction.
We further examine how escalation shape, trajectory-level amplification components, and inference-time strategy refinement affect attack success.
For TBA and SSA, we provide analytical systematization of how distributed intent expands the required observation scope to full sessions, cross-branch linkage, or cross-session linkage.
Together, these analyses motivate multi-turn-aware evaluation protocols and safety architectures that align defenses with the active intent-organization dimensions and the level at which harmful intent becomes observable.

We summarize our contributions as follows:
\begin{itemize}
    \item \textbf{Intent-oriented taxonomy of multi-turn jailbreaks.}
    We argue that the defining characteristic of multi-turn jailbreaks is not the specific prompting technique employed at each step, but how adversaries organize harmful intent. 
    Guided by this perspective, we develop an intent-oriented systematization around four non-exclusive dimensions, providing a unified view of existing multi-turn jailbreak methods and clarifying the relationships among previously disconnected attack designs.

    \item \textbf{Mechanism analysis of single-trajectory attacks.}
    Beyond systematizing prior work, we conduct targeted empirical studies on representative DEA and SRA mechanisms to identify the factors that drive their effectiveness.
    Through controlled ablations, we examine the roles of context accumulation, query budget, escalation structure, strategic adaptation, and selected trajectory-level amplification mechanisms.
    \item \textbf{Implications for evaluation and defense.}
    We further examine how different intent-organization dimensions affect the observability of harmful intent and the corresponding observation scope available to safety mechanisms.
    Our analysis reveals that the required observation scope can extend from turn-level to session-level, cross-branch, or cross-session analysis, exposing a structural mismatch between advanced multi-turn jailbreaks and existing turn-centric defenses.
    Based on these findings, we identify key limitations in current evaluation protocols, benchmark design, and defense strategies, and outline directions toward interaction-aware safety evaluation and robust defenses against adaptive long-horizon attacks.
\end{itemize}

\section{Modeling Multi-Turn Jailbreaks: Formulation and Taxonomy}
\begin{figure*}[t]
    \centering
    \includegraphics[width=0.95\textwidth]{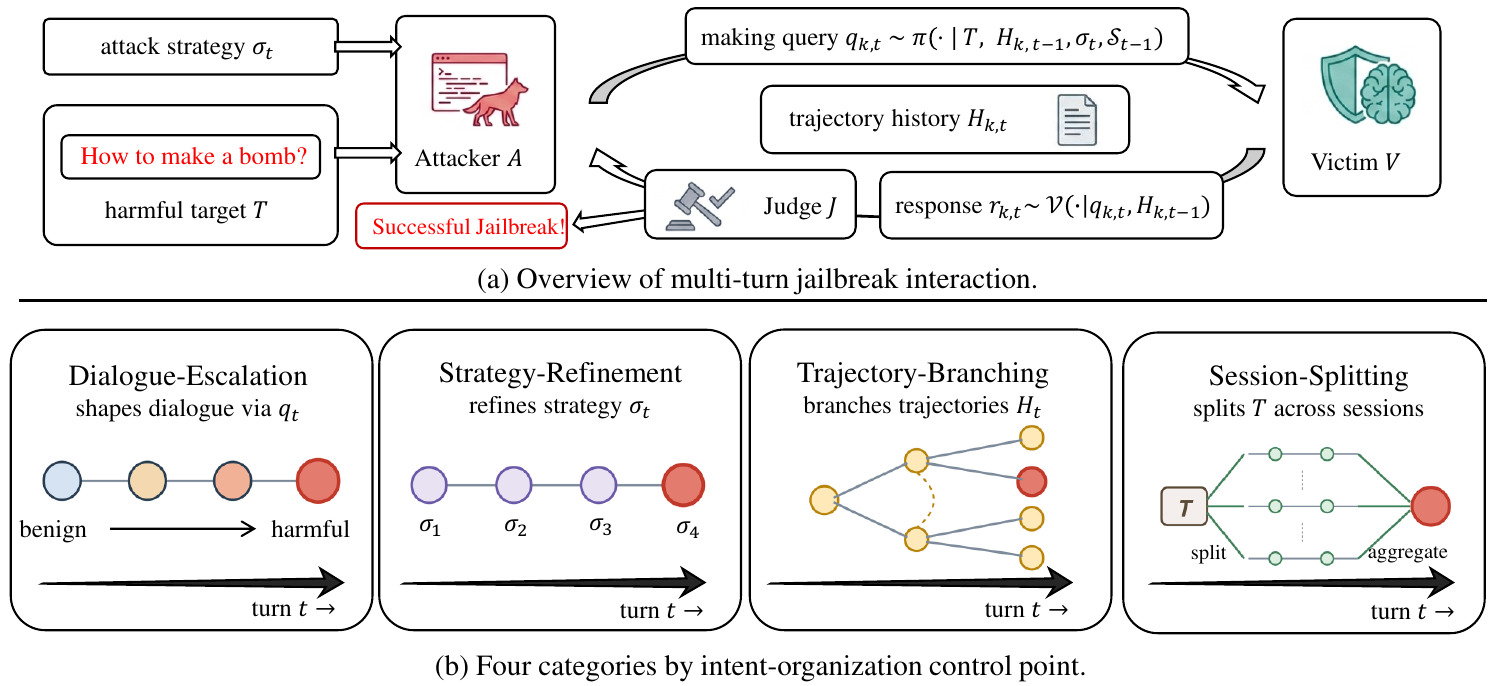}
    \caption{Overview of multi-turn jailbreak. Panel (a) illustrates the interaction loop among the attacker $\mathcal{A}$, victim $\mathcal{V}$, and judge $\mathcal{J}$, with the trajectory history $\mathcal{H}_{k,t}$ updated each turn until the judge signals a successful jailbreak.
    Panel (b) summarizes four non-exclusive intent-organization dimensions along the turn axis, distinguished by the objects through which harmful intent may be organized, namely the query content $q_t$, the strategy trajectory $\sigma_t$, the branching of $\mathcal{H}_t$, and the decomposition of $T$ across sessions.}
    \label{fig:mtj_overview}
\end{figure*}

This section establishes the modeling basis for our intent-oriented analysis of multi-turn jailbreaks. 
We first formalize the interaction process among the attacker, victim, and judge in Subsection~\ref{sec:formulation}. 
We then specify the threat setting in terms of attacker capabilities and defender observation in Subsection~\ref{sec:threat_model}. 
Finally, we derive our taxonomy from where harmful intent is organized across the interaction in Subsection~\ref{sec:taxonomy}. 
\autoref{fig:mtj_overview} provides a visual overview of the interaction loop and the resulting taxonomy.

\subsection{Problem Definition of Multi-Turn Jailbreak}
\label{sec:formulation}

\noindent\textbf{Process.}
We model multi-turn jailbreak as an interactive process in which an adversary advances a harmful intent, encoded as a target objective $T$, across turns, eliciting policy-violating outputs from a victim model through sequential context construction and adaptive querying.
Let $N$ denote the maximum turn budget, with the total number of victim queries bounded by $\sum_{t=1}^{N} K_t$.

\noindent\textbf{Roles.}
The process involves three formal roles:
\begin{itemize}
    \item An attacker $\mathcal{A}$ that implements two conditional distributions, a strategy update $\mu$ and a query policy $\pi$, governing how intent is organized and how queries are generated;
    \item A victim $\mathcal{V}$ that generates responses conditioned on the dialogue history;
    \item A judge $\mathcal{J}$ that scores a candidate output against the target objective $T$ for evaluation.
\end{itemize}
Parameter subscripts (e.g., $\mathcal{V}_\theta$, $\mathcal{J}_\theta$) are reserved for specific instantiations and omitted at the formulation level for uniformity.

\noindent\textbf{Attacker state.}
Across turns, the attacker maintains three internal objects:
\begin{itemize}
    \item A strategy variable $\sigma_t$ that encodes the current attack strategy;
    \item A trajectory set $\mathcal{H}_t = \{H_{k,t}\}_{k=1}^{K_t}$ recording all query-response pairs up to turn $t$ along each trajectory $k$;
    \item A cross-trajectory shared state $\mathcal{S}_t$ that links concurrent trajectories.
\end{itemize}
The cardinality $K_t$ may be fixed throughout the interaction or evolve under $\sigma_t$, which encodes any branching or pruning over trajectories.
When $K_t = 1$, the attack proceeds along a single dialogue thread and $\mathcal{S}_t$ degenerates to an empty placeholder; when $K_t > 1$, multiple trajectories run concurrently and $\mathcal{S}_t$ carries whatever information links them, with the role of this linkage determined by the attack design.

\noindent\textbf{Initialization.}
Each trajectory is initialized with a prior context $C_0$ that captures any system prompt or other initial conditioning ($C_0 = \emptyset$ when none is present), so that $H_{k,0} = C_0$; the initial strategy $\sigma_0$ and shared state $\mathcal{S}_0$ are specified per attack design.

\noindent\textbf{Interaction loop.}
At each turn $t \in \{1, \ldots, N\}$, the attacker first updates its strategy according to the strategy update distribution $\mu$:
\begin{equation}
\label{eq:strategy_update}
    \sigma_t \sim \mu(\cdot \mid \sigma_{t-1},\; \mathcal{H}_{t-1},\; \mathcal{S}_{t-1}).
\end{equation}
The strategy $\sigma_t$ is shared across all active trajectories within a turn; per-trajectory variation arises from the differing $H_{k,t-1}$ and the trajectory-specific use of $\mathcal{S}_{t-1}$ within $\pi$.
Conditioned on this strategy, the attacker generates, for each active trajectory $k$, a query $q_{k,t}$ via the query policy $\pi$:
\begin{equation}
\label{eq:attack_policy}
    q_{k,t} \sim \pi(\cdot \mid T,\; H_{k,t-1},\; \sigma_t,\; \mathcal{S}_{t-1}).
\end{equation}
The victim processes each query together with its trajectory history to produce a response $r_{k,t}$, and the history is updated accordingly:
\begin{equation}
\begin{aligned}
    r_{k,t} &\sim \mathcal{V}(\cdot \mid q_{k,t},\; H_{k,t-1}), \\
    H_{k,t} &= H_{k,t-1} \cup \{(q_{k,t},\, r_{k,t})\}.
\end{aligned}
\label{eq:victim_response}
\end{equation}
The shared state is then refreshed via an update function $\Phi$:
\begin{equation}
\label{eq:state_update}
    \mathcal{S}_t = \Phi(\mathcal{S}_{t-1},\; \mathcal{H}_t).
\end{equation}
Finally, a candidate output is extracted from the trajectory set via an extraction function $\Omega$ and scored by the judge:
\begin{equation}
\label{eq:outcome_extraction}
    O_t = \Omega(\mathcal{H}_t),
    \qquad
    s_t = \mathcal{J}(O_t,\; T).
\end{equation}
In the simplest case, $\Omega$ returns the most recent victim response along the trajectory; more elaborate designs aggregate information across trajectories before producing $O_t$.

\noindent\textbf{Success criterion.}
A multi-turn jailbreak succeeds if the judge score reaches a predefined threshold $\tau$ within the turn budget:
\begin{equation}
\label{eq:success}
    \exists\, t \leq N \quad \text{s.t.} \quad s_t \geq \tau.
\end{equation}
Equivalently, the attack succeeds at the first turn $t^*$ with $s_{t^*} \geq \tau$, and fails if no such $t^*$ exists within the budget.

\noindent\textbf{Generality.}
Algorithm~\ref{alg:general} summarizes this loop.
It captures the core mechanics shared by every multi-turn attack, while Subsection~\ref{sec:threat_model} specifies the capability and observation assumptions under which the loop is instantiated.
The four dimensions of Subsection~\ref{sec:taxonomy} are projections of this shared procedure rather than separate algorithms or mutually exclusive specializations.
A method may activate multiple control points among $\{\mu, \pi, \mathcal{H}_t, \mathcal{S}_t, \Phi, \Omega\}$, and Subsection~\ref{sec:taxonomy} describes the corresponding intent-organization dimensions.

\begin{algorithm}[!t]
    \small
    \caption{General formulation of multi-turn jailbreak. }
    \label{alg:general}
    \begin{algorithmic}[1]
    \REQUIRE target $T$, victim $\mathcal{V}$, judge $\mathcal{J}$, threshold $\tau$, turn budget $N$, prior context $\mathcal{C}_0$, initial strategy $\sigma_0$, shared state $\mathcal{S}_0$, cardinality $K_0$
    \STATE $H_{k,0} \leftarrow \mathcal{C}_0$ \textbf{for} $k = 1, \ldots, K_0$
    \FOR{$t = 1$ \textbf{to} $N$}
        \STATE $\sigma_t \sim \mu(\cdot \mid \sigma_{t-1}, \mathcal{H}_{t-1}, \mathcal{S}_{t-1})$ \COMMENT{may reshape $K_t$ and branching}
        \FOR{$k = 1$ \textbf{to} $K_t$ \textbf{in parallel}}
            \STATE $q_{k,t} \sim \pi(\cdot \mid T, H_{k,t-1}, \sigma_t, \mathcal{S}_{t-1})$
            \STATE $r_{k,t} \sim \mathcal{V}(\cdot \mid q_{k,t}, H_{k,t-1})$
            \STATE $H_{k,t} \leftarrow H_{k,t-1} \cup \{(q_{k,t}, r_{k,t})\}$
        \ENDFOR
        \STATE $\mathcal{S}_t \leftarrow \Phi(\mathcal{S}_{t-1}, \mathcal{H}_t)$ \COMMENT{trivial: $\mathcal{S}_t \equiv \emptyset$ when $K_t \equiv 1$}
        \STATE $O_t \leftarrow \Omega(\mathcal{H}_t)$ \COMMENT{aggregates across trajectories when $K_t > 1$}
        \IF{$\mathcal{J}(O_t, T) \geq \tau$}
            \STATE \textbf{return} Success, $\mathcal{H}_t$
        \ENDIF
    \ENDFOR
    \STATE \textbf{return} Failure, $\mathcal{H}_N$
    \end{algorithmic}
\end{algorithm}

\subsection{Threat Model}
\label{sec:threat_model}

\noindent\textbf{Scope.}
We specify a threat-model space rather than a single fixed setting, because existing multi-turn jailbreak methods differ in attacker access, session organization, and defender observation.
The adversary is given a harmful target objective $T$ and interacts with a deployed LLM or LLM-based system through its conversation interface.
We do not consider attacks that modify the victim model, its safety training data, or the deployment infrastructure.

\noindent\textbf{Attacker capabilities.}
The attacker can issue user-side queries, observe victim responses, and adapt later queries based on earlier interaction outcomes.
Depending on the method, the attacker may be a human operator, a fixed template, an auxiliary LLM, an optimizer, or a multi-agent system.
We distinguish black-box access, where only input and output behavior is visible, from white-box access, where internal signals such as attention scores or hidden states may also be used.
Unless a method explicitly relies on internal model signals, we treat it as operating under black-box interaction.

\noindent\textbf{Session structure.}
We consider three session settings:
\begin{itemize}
    \item \textbf{Single-session attacks} proceed within one continuous dialogue.
    \item \textbf{Shared-state attacks} explore multiple trajectories for the same target and share information across branches.
    \item \textbf{Isolated cross-session attacks} distribute sub-objectives across independent sessions and recombine their outputs outside any single session.
\end{itemize}

\noindent\textbf{Defender observation.}
The deployment-time monitor $\mathcal{M}$ is distinct from the evaluation judge $\mathcal{J}$.
The judge $\mathcal{J}$ is used to measure whether the extracted output satisfies $T$, whereas $\mathcal{M}$ represents the safety mechanism that the deployed system can actually apply.
We distinguish three observation scopes for $\mathcal{M}$:
\begin{itemize}
    \item \textbf{Turn-level}, where only the current query, response, or local turn context is inspected;
    \item \textbf{Session-level}, where the full dialogue history of one session is inspected;
    \item \textbf{Cross-session}, an umbrella scope in which multiple sessions or trajectories, including related branches, can be linked and analyzed jointly.
\end{itemize}
The observation scope of $\mathcal{M}$ determines where harmful intent becomes detectable: some attacks only reveal intent at the session level, while session-splitting attacks require cross-session linkage.

\subsection{Taxonomy based on Intent Analysis}
\label{sec:taxonomy}
\noindent\textbf{Intent-organization dimensions.}
Surveying existing multi-turn methods, we identify recurring control points through which adversaries organize harmful intent over an interaction.
These control points correspond to formal objects from Subsection~\ref{sec:formulation}, and multiple control points may be active in the same method.
As shown in~\autoref{tbl:intent_control}, the taxonomy contains four top-level dimensions:
\begin{itemize}
    \item \textbf{Dialogue-escalation Attacks (DEA)} organize intent by progressively shaping the dialogue context through query content $q$ and policy $\pi$.
    \item \textbf{Strategy-refinement Attacks (SRA)} keep the harmful target fixed and refine the strategy trajectory $\sigma$ via update $\mu$.
    \item \textbf{Trajectory-branching Attacks (TBA)} organize intent through branching histories $\mathcal{H}_t$ by activating multiple trajectories $K_t$.
    \item \textbf{Session-splitting Attacks (SSA)} split the target $T$ across isolated sessions and recombine outputs through $\Omega$.
\end{itemize}
These four dimensions capture the cross-turn control points observed across the surveyed methods.
At the method level, membership is non-exclusive and a method receives every top-level label supported by its constitutive attack process.

\noindent\textbf{Specializations of the general loop.}
As summarized in~\autoref{tbl:intent_control}, each dimension describes a canonical specialization of Algorithm~\autoref{alg:general} centered on a particular intent-organization control point.
DEA and SRA can co-occur in a single trajectory when query construction and strategy refinement both contribute constitutively to the attack.
TBA can also incorporate strategy refinement through branch feedback, while SSA can contain DEA or SRA mechanisms within each isolated session.
Multi-label assignments reflect a control point's substantive role in constructing or executing the attack, rather than the incidental use of a related technique. For example, a method is labeled TBA only when information from multiple trajectories guides exploration, selection, or later attack decisions; generating several independent candidate prompts without such coordination does not by itself establish TBA membership.
We develop each dimension in Subsection~\ref{sec:pe_mtj} through Subsection~\ref{sec:ss_mtj}.

\begin{table}[t]
    \centering
    \footnotesize
    \setlength{\tabcolsep}{4.5pt}
    \renewcommand{\arraystretch}{1.4} 
    \setlength{\aboverulesep}{0pt}    
    \setlength{\belowrulesep}{0pt}  
    \caption{Intent-organization dimensions in the multi-label taxonomy. Each dimension highlights a primary object, shown in \colorbox{ctrlgreen}{green}, through which harmful intent is organized. The objects shown in \colorbox{fixedred}{red} are held fixed or trivialized only in the canonical standalone form; they may also be active when dimensions co-occur.}
    \label{tbl:intent_control}
    \resizebox{\columnwidth}{!}{%

    \begin{tabular}{@{}>{\centering\arraybackslash}m{0.18\columnwidth} >{\columncolor{ctrlgreen}}l >{\columncolor{fixedred}[\tabcolsep][0pt]}l@{}}
        \toprule
        \textbf{Dimension} & \textbf{Primary intent-organization object} & \textbf{Other objects in canonical form} \\
        \midrule
        DEA & query content $q$ (policy $\pi$)      & $\mu$ fixed; $K_t\!\equiv\!1$ \\
        SRA & strategy $\sigma$ (update $\mu$)       & $T$ fixed; $K_t\!\equiv\!1$ \\
        TBA & branching of $\mathcal{H}_t$ (via $K_t$) & $T$ kept whole; $\Phi$ aggregates \\
        SSA & decomposition of $T$                    & split sessions; recombine in $\Omega$ \\
        \bottomrule
    \end{tabular}%
    }
\end{table}

\subsection{Comparison with Existing Works}
Existing work has studied LLM jailbreaks through broad security and privacy surveys, jailbreak-focused taxonomies, empirical benchmarks, frameworks, and recent SoK studies. 
Broad surveys usually discuss jailbreak as part of the general LLM security landscape~\cite{shayegani2023llm_vulnerabilities_survey, yao2024llm_security_privacy, das2025llm_security_privacy_survey,yi2024jailbreak}. 
Jailbreak-focused works organize attacks by attacker access, defense placement, prompt techniques, or attack-defense performance~\cite{yi2024jailbreaksurvey, xu2024comprehensive, knowlton2026prompt}, while other studies decompose prompt security threats~\cite{hong2025sok}, extend the scope to LLMs, MLLMs, and agents~\cite{mao2025llms, ying2026safebench, liu2025agentsafe, liang2025safemobile}, or systematize jailbreak vectors~\cite{hakim2026jailbreaking}. 
Recent SoK studies further examine jailbreak guardrails~\cite{wang2025sok} and jailbreak robustness under multi-dimensional evaluation settings~\cite{xu2026sok}. 
These studies provide valuable coverage of jailbreak attacks, defenses, and evaluation protocols, but they still leave two critical gaps for multi-turn jailbreaks.

\noindent\textbf{\Circled{1} Limited treatment of multi-turn interaction.}
Existing works rarely treat multi-turn jailbreak as an independent threat model. 
When multi-turn attacks are covered, they are usually described as a subtype of prompt-based jailbreak or as an extension of single-turn attacks. 
This view overlooks that harmful intent may not appear in any single prompt, but may emerge through dialogue construction, response-based adaptation, or cross-session decomposition. 
As a result, prior work does not fully explain how interaction changes the attack surface or where harmful intent becomes observable to a defender.

\noindent\textbf{\Circled{2} Limited mechanism-level analysis.}
Existing evaluations mainly report outcome metrics such as attack success rate, defense success rate, utility preservation, or judge agreement. 
These metrics compare attacks and defenses, but they do not reveal the logic that makes multi-turn jailbreaks effective. 
In particular, it remains unclear whether success is driven by accumulated context, query count, intent ordering, strategy trajectory, or objective decomposition across branches and sessions. 
This makes it difficult to determine whether a defense fails because of weak classification, insufficient observation scope, or an evaluation protocol that is not interaction-aware.

Our work addresses these gaps by centering multi-turn jailbreak as an interaction-level threat and classifying attacks according to how harmful intent is organized across turns, trajectories, and sessions. 
\autoref{tab:survey_comparison} summarizes how this intent-oriented perspective differs from existing surveys, benchmarks, frameworks, and SoK studies.

\begin{table*}[t]
    \centering
    \caption{Comparison with existing surveys, benchmarks, frameworks, and SoKs on LLM jailbreaks from an intent-oriented interaction perspective.
    \textbf{MT}: whether multi-turn jailbreak is treated as a first-class threat.
    \textbf{Interaction Unit}: the primary unit of analysis used by the work.
    \textbf{Intent Analysis}: whether the work analyzes how harmful intent is organized, advanced, or hidden across turns, trajectories, or sessions.
    \textbf{Mechanism}: whether targeted empirical analyses or controlled ablations are conducted to explain attack mechanisms.
    \textbf{Detect Surface}: whether the work studies the level at which harmful intent becomes observable.
    \textbf{Evaluation}: whether evaluation protocols, judges, or judge reliability are systematically discussed.
    \textbf{Defense}: whether defenses or guardrails are systematically discussed.
    \full\ = covered, \none\ = not covered, \half\ = partially covered.}
    \label{tab:survey_comparison}
    \footnotesize
    \resizebox{\textwidth}{!}{%
    \begin{tabular}{l c c p{4.2cm} c p{2.5cm} c c c c c}
        \toprule
        \multirow{2}{*}{\textbf{Work}}
        & \multirow{2}{*}{\textbf{Year}}
        & \multirow{2}{*}{\textbf{Type}}
        & \multirow{2}{*}{\textbf{Primary Axis}}
        & \multicolumn{3}{c}{\textbf{Interaction Perspective}}
        & \multicolumn{2}{c}{\textbf{Mechanism Perspective}}
        & \multicolumn{2}{c}{\textbf{Evaluation / Defense Perspective}} \\
        \cmidrule(lr){5-7} \cmidrule(lr){8-9} \cmidrule(lr){10-11}
        & & &
        & \textbf{MT}
        & \textbf{Interaction Unit}
        & \textbf{Intent Analysis}
        & \textbf{Mechanism}
        & \textbf{Detect Surface}
        & \textbf{Evaluation}
        & \textbf{Defense} \\
        \midrule

        \cite{shayegani2023llm_vulnerabilities_survey}
          & 2023
          & Survey
          & Adversarial attack surface
          & \none
          & Prompt / model
          & \none
          & \none
          & \none
          & \none
          & \half \\

        \cite{yi2024jailbreaksurvey}
          & 2024
          & Survey
          & Attacker access \& defense placement
          & \half
          & Prompt / turn
          & \none
          & \none
          & \half
          & \full
          & \full \\

        \cite{xu2024comprehensive}
          & 2024
          & Benchmark
          & Attack defense comparison
          & \none
          & Prompt / instance
          & \none
          & \half
          & \none
          & \full
          & \full \\

        \cite{yao2024llm_security_privacy}
          & 2024
          & Survey
          & Security \& privacy lifecycle
          & \none
          & System lifecycle
          & \none
          & \none
          & \half
          & \half
          & \full \\

        \cite{das2025llm_security_privacy_survey}
          & 2025
          & Survey
          & Security \& privacy lifecycle
          & \none
          & System lifecycle
          & \none
          & \none
          & \half
          & \half
          & \full \\

        \cite{mao2025llms}
          & 2025
          & Survey
          & Technology evolution 
          & \half
          & LLM / MLLM / agent
          & \half
          & \none
          & \half
          & \full
          & \full \\

        \cite{hong2025sok}
          & 2025
          & SoK
          & Prompt-security threat model
          & \half
          & Prompt / threat unit
          & \none
          & \half
          & \half
          & \full
          & \full \\

        \cite{knowlton2026prompt}
          & 2026
          & Survey
          & Prompt-based technique taxonomy
          & \half
          & Prompt / turn
          & \none
          & \half
          & \none
          & \full
          & \full \\

        \cite{hakim2026jailbreaking}
          & 2026
          & Survey
          & Jailbreak vector taxonomy
          & \half
          & Attack vector
          & \half
          & \none
          & \half
          & \full
          & \full \\

        \cite{wang2025sok}
          & 2026
          & SoK
          & Jailbreak guardrail taxonomy
          & \half
          & Guardrail
          & \none
          & \half
          & \half
          & \full
          & \full \\

        \cite{xu2026sok}
          & 2026
          & SoK
          &  robustness evaluation
          & \half
          & Security Cube
          & \none
          & \half
          & \half
          & \full
          & \full \\

        \midrule

        \textbf{Ours}
          & 2026
          & SoK
          & \textbf{Intent organization}
          & \full
          & \textbf{Interaction path}
          & \full
          & \full
          & \full
          & \full
          & \half \\

        \bottomrule
    \end{tabular}%
    }
\end{table*}

\section{Research Methodology}
\label{sec:literature_collection}

We conducted a structured literature review following the principles of systematic literature reviews (SLRs)~\cite{Kitchenham2015} and the guidance for combining peer-reviewed and other relevant sources in multivocal reviews~\cite{garousi2019grey}. The review was designed to support both the construction of our taxonomy and the comparison with existing surveys, benchmarks, frameworks, and SoK studies. The final search update was completed on August 10, 2026.

\noindent\textbf{Review question and scope.} Our overarching question is: \emph{How do existing studies characterize, implement, and evaluate multi-turn jailbreak attacks, and how can these attacks be systematized according to the organization of harmful intent?} We define a core multi-turn jailbreak method as an attack involving at least two attacker queries and their corresponding target-model responses, or one in which the attacker coordinates multiple trajectories or sessions toward a jailbreak outcome. This scope includes sequential dialogue attacks, branching interaction processes, and attacks distributed across sessions. It excludes single prompts containing fabricated dialogues, attacks executed as single-turn requests after offline preprocessing, and adjacent memory or prompt-injection studies without an explicit jailbreak outcome.

\noindent\textbf{Development of the search strategy.} Before executing the formal search, we reviewed representative multi-turn jailbreak papers and related reporting studies to identify how the literature describes interaction length, adaptive querying, trajectory search, decomposition, memory, multimodality, and tool use. These observations were used to construct query families rather than to define the final corpus in advance. The resulting queries combined interaction terms (``multi-turn,'' ``multi-round,'' ``conversational,'' ``iterative,'' and ``adaptive''), attack terms (``jailbreak,'' ``red teaming,'' and ``safety bypass''), system terms (``LLM,'' ``MLLM,'' ``language model,'' and ``LLM agent''), and mechanism terms including branching, tree search, decomposition, cross-session, memory, multimodal, tool use, and artifact.

\noindent\textbf{Source selection and search.} We searched arXiv, OpenReview, ACL Anthology, public USENIX Security, ACM Digital Library, and IEEE Xplore pages, together with official project or paper pages when available. We also checked relevant security, machine learning, and natural language processing venues and traced backward and forward citations from the retrieved literature. These sources were selected to cover both peer-reviewed publications and recent preprints or project releases, which are particularly important in the rapidly developing jailbreak literature. The search log records the source, query family, search date, accessible records screened, and candidate studies recovered.

\noindent\textbf{Eligibility, exclusion, and relevance checking.} Search results were merged and deduplicated using titles, persistent identifiers, publication versions, and method aliases. We then applied the eligibility criteria to the title, abstract, and available full text. Records were excluded when they did not contain an explicit multi-turn jailbreak outcome, reduced the attack to a single-turn request, or only used a synthetic dialogue as part of one prompt. Studies focused on adjacent memory integrity or prompt-injection problems were retained only when they explicitly evaluated a jailbreak consequence. The search produced 60 unique records. After excluding four records that failed these criteria, 56 records remained: 45 core multi-turn jailbreak methods and 11 surveys, benchmarks, frameworks, or SoK studies. The latter studies provide the comparison basis summarized in Table~\ref{tab:survey_comparison}.

\noindent\textbf{Extraction and synthesis.} For each retained record, we extracted bibliographic metadata, provenance, interaction structure, attacker resources, modality, tool or agent use, observation scope, evaluated targets, datasets, metrics, code availability, and review basis.
For reporting studies, we additionally recorded their primary scope, interaction unit, treatment of intent organization, mechanism analysis, observation scope, evaluation perspective, and defense coverage.
For each core method, we coded every applicable intent-organization dimension according to the definitions established in Section~\ref{sec:taxonomy}.
Category membership was non-exclusive, while DEA, SRA, TBA, and SSA sub-paradigms were recorded separately within their applicable top-level dimensions.
Multimodality, tool use, memory, optimization, artifact persistence, and human participation were coded as orthogonal attributes rather than as additional intent-organization dimensions.
    The 45 core methods therefore represent 45 unique methods, not a partition obtained by summing category memberships.
When category counts are reported, they denote category-membership counts and should not be interpreted as a partition of the 45 methods.
    A second reviewer checked the retained attack and reporting records and reviewed the eligibility and multi-label coding assignments. The reviewers independently agreed on 40 of the 45 core-method assignments; they discussed and reconciled the remaining five cases before finalizing the coding.
    To make these judgments reproducible, coders triangulated evidence from each paper's method description, execution workflow, and evaluation setup. Disagreements primarily concerned whether a reported strategy update was an online part of the attack process or merely offline preprocessing, and whether multiple branches required defender-side linkage to constitute coordinated trajectory exploration. The reviewers resolved these cases by returning to the primary sources and applying the definitions in Section~\ref{sec:taxonomy} consistently.

\section{Taxonomy-Guided Analysis of Multi-Turn Jailbreak Methods}

This section uses the taxonomy above to organize existing multi-turn jailbreak methods according to how they organize harmful intent across turns.
\autoref{tbl:mtj_taxonomy} presents representative mechanism rows, whereas \autoref{app:method-level-taxonomy} records the complete method-level taxonomy with multi-label assignments.
We therefore discuss each dimension through its organizing mechanism, representative sub-paradigms, and associated observation scope rather than treating every method as a separate narrative thread.

\subsection{Dialogue-escalation Attacks}
\label{sec:pe_mtj}

Dialogue-escalation Attacks (DEA) are characterized by organizing harmful intent through query content and dialogue context.
In a canonical DEA instantiation, an attacker follows a static or semi-static interaction plan, using benign, ambiguous, or weakly related turns to make a later prohibited request appear coherent, justified, and contextually licensed~\cite{yu2024don, deng2023masterkey}.
This specialization can be described by a single trajectory ($K_t \equiv 1$) with no shared branch state ($\mathcal{S}_t \equiv \emptyset$), but neither condition is necessary for DEA membership.
Its central control point is the query policy $\pi$, which progressively embeds harmful intent into dialogue history.

This design exploits model-level vulnerabilities such as attention allocation, recency bias, instruction drift, and conversational consistency~\cite{zhou2024alignment, arditi2024refusal, laban2025llms}, as well as interaction-level compliance effects such as Foot-in-the-Door and semantic consistency~\cite{weng2025foot}. 
The central mechanism is not that the attacker asks more questions, but that earlier turns reshape how later turns are interpreted. 
DEA manipulates how dialogue history frames intent and constraints, allowing later requests to inherit the apparent legitimacy of earlier turns~\cite{wei2023jailbroken, liu2024making}. 
We organize DEA into three sub-paradigms: cognitive and psychological strategies, context manipulation and attention shifting, and task decomposition and knowledge aggregation.

\noindent\textbf{Cognitive Biases and Psychological Strategies.}
This sub-paradigm treats multi-turn interaction as a gradual compliance-shaping process. 
The attacker first establishes agreement on benign or weakly related requests, then escalates toward the harmful target while preserving conversational continuity. 
Crescendo~\cite{russinovich2025great}, FITD~\cite{weng2025foot}, and Echo Chamber~\cite{alobaid2026echo} implement this mechanism through gradual escalation, compliance momentum, and semantic continuity, respectively.
Each keeps individual turns locally plausible while the overall trajectory moves toward a prohibited endpoint.
Thus, harmful intent need not be explicit at the outset because it becomes progressively legible through the staged path while remaining less salient to turn-local checks.
Existing studies rarely separate the effect of staged organization from that of longer context.

\noindent\textbf{Context Manipulation and Attention Shifting.}
A second line of DEA methods shapes the context used to interpret a later request.
CIA~\cite{Cheng2024} and CFA~\cite{Sun2024} groom dialogue context and exploit instructional drift or recency; AMA~\cite{wuanalogy} uses analogy to transfer a benign response structure to harmful content; and ActorAttack~\cite{ren2024derail} automates contextually aligned attack construction.
Chain of Attack~\cite{yang2025chain} and GRAF~\cite{tang2025graf} further combine context construction with explicit strategy refinement.
Across these implementations, the shared pattern is dialogue-context grooming, analogy or instructional drift, and attention shifting, rather than repetition alone.
Their observation scope is therefore often trajectory-level.

\begin{table*}[!ht]
\centering
\scriptsize
\caption[Representative intent-organization dimensions and sub-paradigms.]{%
Representative intent-organization dimensions and sub-paradigms.
Repeated references denote multi-label membership, with complete method-level coding in Appendix~\autoref{app:method-level-taxonomy}.
\textsuperscript{1}\,\textbf{Attack Mechanism}: \textbf{Template}, predetermined workflow without inference-time strategy adaptation; \textbf{LLM}, adaptive attacker LLM; \textbf{RL}, reinforcement learning; \textbf{MA}, multiple attacker agents; \textbf{Human}, human red teamer.
\textsuperscript{2}\,\textbf{Interaction Scope}: \textbf{Single}, one trajectory; \textbf{Multi}, parallel trajectories or sessions.
\textsuperscript{3}\,\textbf{Session Continuity}: \textbf{Linear}, one thread; \textbf{Branching}, parallel paths; \textbf{Fragmented}, isolated sessions.
\textsuperscript{4}\,\textbf{Observation Scope}: \textbf{Turn}, one message; \textbf{Session}, one full session; \textbf{Cross-session}, linked sessions or trajectories.}
\label{tbl:mtj_taxonomy}

\begin{tblr}{
  width   = \linewidth,
  colspec = {|X[1.4]|X[1.5]|X[1.2]|X[0.9]|X[1.1]|X[1.2]|X[1.0]|},
  cells = {halign=c,valign=m},
  colsep  = 3pt,
  rowsep  = 0.5pt,
  hlines, vlines,
  row{1}  = {font=\bfseries},
}

\textbf{Intent Dimension} &
\textbf{Sub-paradigm} &
\textbf{Attack Mechanism\textsuperscript{1}} &
\textbf{Interaction Scope\textsuperscript{2}} &
\textbf{Session Continuity\textsuperscript{3}} &
\textbf{Observation Scope\textsuperscript{4}} &
\textbf{References} \\

\SetCell[r=10]{m} Dialogue-escalation Attacks (DEA)
  & \SetCell[r=2]{m} Psychological
    & Template & Single & Linear & Session-level & \cite{weng2025foot} \\
  & & LLM      & Single & Linear & Session-level & \cite{russinovich2025great,alobaid2026echo} \\
  & \SetCell[r=6]{m} Context Manipulation
    & Template & Single & Linear & Session-level & \cite{Sun2024,wuanalogy,MRCJ,li2026coopguard,kulshreshtha2026multi} \\
  & & LLM      & Single & Linear & Session-level & \cite{Cheng2024,ren2024derail,Ying2025,lin2026icon,du2025multi,pavlova2024automated,chen2025strategize,mo2025redcoder,yang2025chain,nihal2026pattern,li2026knowledge,das2026multi,choi2026multi,tang2025graf,bhuiya2026plague} \\
  & & RL       & Single & Linear & Session-level & \cite{ramesh2025efficient,wang2026jailbreaking,feng2026sema} \\
  & & MA       & Single & Linear & Session-level & \cite{rahman2025x,Wahreus2025Prompt,Srivastav2025Safe,li2025stac} \\
  & & Human    & Single & Linear & Session-level & \cite{li2024llm} \\
  & & MA       & Multi  & Branching & Cross-session & \cite{rafieiasl2025nexus,narula2025harmnet} \\
  & \SetCell[r=2]{m} Task Decomposition
    & Template & Single & Linear & Session-level & \cite{Liu2024,yang2025jigsaw,zhao2026when} \\
  & & LLM      & Single & Linear & Session-level & \cite{Zhou2024,Wang2024,Zhao2025} \\

\SetCell[r=10]{m} Strategy-refinement Attacks (SRA)
  & Fixed-strategy
    & Template & Single & Linear & Turn    & \cite{MRCJ,li2026coopguard,kulshreshtha2026multi} \\
  & \SetCell[r=6]{m} Adaptive-optimization
    & LLM      & Single & Linear & Turn    & \cite{du2025multi,zhou2025tempest,russinovich2025great,ren2024derail,Ying2025,lin2026icon,Zhao2025} \\
  & & RL       & Single & Linear & Turn    & \cite{ramesh2025efficient,wang2026jailbreaking,feng2026sema} \\
  & & LLM      & Single & Linear & Session-level & \cite{feng2026jail,choi2026multi} \\
  & & RL       & Single & Linear & Session-level & \cite{xiong2026trojail} \\
  & & RL       & Multi  & Branching & Cross-session & \cite{anonymous2026multiturn} \\
  & & Template & Multi  & Branching & Cross-session & \cite{liu2025let} \\
  & \SetCell[r=3]{m} Agent-based
    & LLM      & Single & Linear & Session-level & \cite{pavlova2024automated,chen2025strategize,mo2025redcoder,Wang2024,yang2025chain,li2026knowledge,tang2025graf,bhuiya2026plague} \\
  & & MA       & Single & Linear & Session-level & \cite{rahman2025x,Wahreus2025Prompt,li2025stac} \\
  & & Human    & Single & Linear & Session-level & \cite{li2024llm} \\

\SetCell[r=5]{m} Trajectory-branching Attacks (TBA)
  & \SetCell[r=5]{m} /
    & LLM      & Multi  & Branching  & Cross-session & \cite{zhou2025tempest,feng2026jail} \\
  & & Template & Multi  & Branching  & Cross-session & \cite{kulshreshtha2026multi,liu2025let} \\
  & & RL       & Multi  & Branching  & Cross-session & \cite{anonymous2026multiturn} \\
  & & MA       & Multi  & Branching  & Cross-session & \cite{rafieiasl2025nexus,narula2025harmnet} \\
  & & Human    & Multi  & Branching  & Cross-session & \cite{li2024llm} \\

\SetCell[r=2]{m} Session-Splitting Attacks (SSA)
  & \SetCell[r=2]{m} /
    & MA       & Multi  & Fragmented & Cross-session & \cite{Wahreus2025Prompt,Srivastav2025Safe} \\
  & & Template & Multi  & Fragmented & Cross-session & \cite{lin2026context} \\

\end{tblr}
\end{table*}

\noindent\textbf{Task Decomposition and Knowledge Aggregation.}
The third DEA sub-paradigm advances harmful intent by decomposing a prohibited objective into individually benign sub-queries. 
Each turn can appear harmless in isolation, but the sequence collectively supplies the information needed for the harmful objective. 
The attack succeeds when partial outputs are aggregated, either by the attacker or through the model's own reasoning trajectory. 
Compared with psychological escalation and context manipulation, this sub-paradigm hides intent by distributing the target across apparently disconnected local tasks.

Speak-Out-of-Turn~\cite{Zhou2024}, MRJ-Agent~\cite{Wang2024}, and Imposter.AI~\cite{Liu2024} split objectives into covert or professionally motivated sub-queries; SIREN~\cite{Zhao2025} learns decomposition strategies, and RACE~\cite{Ying2025} uses guided reasoning to make a harmful conclusion emerge from a benign-looking chain.
Prompt, Divide, and Conquer~\cite{Wahreus2025Prompt} and Context-Fractured Decomposition Attacks~\cite{lin2026context} show that aggregation can also extend beyond one linear session through distributed prompts or artifacts.
DEA therefore organizes intent through both escalation and benign-looking sub-goals.
The relevant observation boundary depends on whether aggregation remains in one dialogue or spans sessions.

\noindent\textbf{Observations across Sub-Paradigms.}
Despite their surface differences, the three sub-paradigms organize earlier turns to reshape the context in which the final request is evaluated, making harmful intent clearest when the trajectory is read as a whole.

This reliance on trajectory-level intent organization, however, co-varies with a factor that existing studies do not isolate.
Methods that organize intent more carefully also produce longer dialogue histories, so it remains unclear whether effectiveness is driven by intent arrangement or by accumulated context.
A related ambiguity appears at the component level: DEA methods often combine additional amplification components, such as diversified semantic pathways and embedded multi-step reasoning chains, without ablation.
Their individual contributions therefore remain unknown.

\begin{takeawaybox}
\label{takeaway:pe}
DEA derives its effectiveness from organizing earlier turns to reshape how later ones are interpreted, often making harmful intent most legible at the trajectory level.
\end{takeawaybox}

The two unresolved factors identified above motivate two questions for 
controlled ablation in Section~\ref{sec:mechanism}.

\begin{rqbox}
\label{q1}
Deliberately organized DEA methods also produce longer dialogues and more queries.
Is effectiveness determined by intent organization across the trajectory or by accumulated multi-turn context?
\end{rqbox}

\begin{rqbox}
\label{q2}
DEA methods may embed amplification components, such as diversified semantic pathways and multi-step reasoning chains.
How much does each component contribute to attack success?
\end{rqbox}

\subsection{Strategy-refinement Attacks}
\label{sec:se_mtj}

Strategy-refinement Attacks (SRA) are characterized by explicit control over the strategy state or its update.
DEA emphasizes how query content organizes intent, whereas SRA emphasizes how $\mu$ selects, escalates, or revises the route.
A common specialization adopts a fixed target and a single trajectory ($K_t \equiv 1$, $\mathcal{S}_t \equiv \emptyset$).
The attacker may follow a pre-designed schedule, optimize choices from feedback, or delegate strategic reasoning to an LLM-based agent.
Accordingly, we organize SRA into three sub-paradigms according to how $\mu$ is realized: Fixed-strategy, Adaptive-optimization, and Agent-based.

\noindent\textbf{Fixed-strategy Jailbreak.}
\label{sec:fixed-strategy}
Fixed-strategy SRA methods construct the entire strategy sequence offline through human design, with no strategy revision at inference time. 
The attacker specifies a staged route in advance and executes it as written, usually increasing pressure or directness while keeping intermediate turns plausible. 
This makes the attack simple and reproducible, but limits recovery once the victim model resists the planned trajectory.

MRCJ~\cite{MRCJ} and EMRA in CoopGuard~\cite{li2026coopguard} exemplify offline schedules and deterministic prompt transformations.
Both stage increasing directness without revising the plan at inference time.
These implementations trade feedback adaptation for reproducibility and low cost, while their staged prompts may simultaneously create DEA-style contextual support.
Because they often increase both strategy organization and dialogue length, their separate contributions remain difficult to isolate.

\noindent\textbf{Adaptive-optimization Jailbreak.}
\label{sec:adaptive-optimization}
Adaptive-optimization SRA methods introduce the inference-time decision channel that fixed strategies lack. 
They cast strategy selection as feedback-driven optimization: the attacker proposes candidate strategies, observes or scores the victim's response, and updates future choices according to an explicit success signal. 
Compared with fixed schedules, this makes the strategy trajectory less rigid, but increases query cost and stochasticity.

ASJA~\cite{du2025multi} uses evolutionary candidate recombination, SoC-MAB~\cite{ramesh2025efficient} uses black-box bandit feedback, and iMIST~\cite{wang2026jailbreaking} learns a reward-driven action policy.
TROJail~\cite{xiong2026trojail} and JAIL~\cite{feng2026jail} extend this family with trajectory-level process rewards and adaptive optimization.
The common mechanism is algorithmic search under a success signal.

\noindent\textbf{Agent-based Jailbreak.}
\label{sec:agent-based}
Agent-based SRA methods realize $\mu$ through an LLM-driven planner. 
Instead of selecting from a fixed schedule or optimizing a scalar reward, the attacker LLM observes the victim's response, reflects on failure modes, and revises the strategy in natural-language terms. 
This supports long-horizon planning that is difficult to express as prompt transformations or reward maximization alone.

GOAT~\cite{pavlova2024automated} revises strategies after failed attempts; GALA~\cite{chen2025strategize} maintains tactical context; RedCoder~\cite{mo2025redcoder} distills reusable code-generation tactics; and X-Teaming~\cite{rahman2025x} distributes planning, execution, verification, and rewriting across agents.
PLAGUE~\cite{bhuiya2026plague}, STAC~\cite{li2025stac}, Chain of Attack~\cite{yang2025chain}, and GRAF~\cite{tang2025graf} illustrate further combinations of tactic memory, tools, decomposition, and context construction.
The relevant mechanism is policy refinement, regardless of whether it is implemented by one attacker, several agents, or a tool chain.

\noindent\textbf{Observations across Sub-Paradigms.}
The three sub-paradigms share a single principle: the attacker plans the strategy route toward a harmful objective, which is often held fixed in canonical implementations.
As in DEA, this principle faces the organization-versus-accumulation confound: strategy sequences that organize intent more deliberately also produce longer dialogue histories, and existing evaluations do not control for one while varying the other.

A second observation arises from Fixed-strategy SRA: MRCJ varies the increment profile within ordered sequences while holding the endpoint malice level and turn count fixed.
This isolates whether the optimal escalation shape depends on the target model, a question not addressed by \textbf{RQ1}.

A third observation concerns the sub-paradigm hierarchy. The three sub-paradigms represent increasing inference-time decision-making, but whether this produces a consistent effectiveness progression has not been tested under unified conditions.
The result may depend on the target model: fixed schedules may suffice in some settings, while adaptive and agent-based methods may help more in others.

\begin{takeawaybox}
\label{takeaway:se}
SRA makes the strategy route an explicit planning object, complementing the content-shaping mechanisms emphasized by DEA.
\end{takeawaybox}

As discussed above, \textbf{RQ1} extends to SRA and will be addressed with evidence from both categories in Section~\ref{sec:mechanism}.
Two further questions are specific to SRA.

\begin{rqbox}
\label{q3}
Given the evidence from \textbf{RQ1}, does the optimal escalation profile depend on the target model, or does a single trajectory shape universally suffice?
\end{rqbox}

\begin{rqbox}
\label{q4}
The three SRA sub-paradigms represent an increasing degree of inference-time decision-making over the strategy trajectory.
Does this progression correspond to a consistent increase in attack effectiveness, and under what conditions does the stronger organization justify its additional cost?
\end{rqbox}

\subsection{Trajectory-branching Attacks}
\label{sec:ps_mtj}
Trajectory-branching Attacks (TBA) are characterized by deliberate branching or coordinated exploration over related trajectories.
A common formalization uses $K_t > 1$ concurrent trajectories and shared state $\mathcal{S}_t$, so that $\Phi$ aggregates cross-branch signals and $\pi$ guides subsequent exploration.
The defining control point is the branching structure of $\mathcal{H}_t$.

TAP~\cite{mehrotra2024tree} provides the canonical tree-search case, expanding and pruning candidate prompts.
Tempest~\cite{zhou2025tempest} uses automated breadth-first exploration and propagates partial-compliance signals, while SLIP~\cite{kulshreshtha2026multi} uses a template-based breadth-first expansion and MHJ~\cite{li2024llm} documents human-guided branching.
NEXUS~\cite{rafieiasl2025nexus}, HarmNet~\cite{narula2025harmnet}, and JAIL~\cite{feng2026jail} further combine network or path exploration with adaptive strategy updates and, in some cases, content manipulation.
Thus, the shared mechanism is coordinated exploration rather than a particular search algorithm.
Its detection implication is cross-session observation, including cross-branch analysis of related histories.
For TBA, this requirement does not necessarily mean that the trajectories originate in independently initiated sessions. Rather, the defender must be able to associate trajectories that share a target, common history, or feedback used for expansion and pruning. Treating these histories separately can hide the search-level pattern even when every individual trajectory is visible.
This analysis must preserve shared context and relations among alternatives; otherwise, each branch can appear ordinary and coordinated exploration is lost. Evaluation should test this process when no individual branch provides sufficient evidence on its own.
The limited experimental coverage in Section~\ref{sec:mechanism} reflects implementation coverage rather than a separation of TBA from DEA or SRA.

\begin{takeawaybox}
\label{takeaway:ps}
In Trajectory branching attacks, harmful intent is organized through coordinated trajectory exploration rather than any individual dialogue, requiring safety mechanisms to reason over the search process instead of isolated conversations.
\end{takeawaybox}

\subsection{Session-Splitting Attacks}
\label{sec:ss_mtj}
Session-Splitting Attacks (SSA) organize harmful intent by decomposing a malicious objective across isolated sessions or agentic sub-tasks, so that each session receives only a benign-looking partial objective and prohibited content emerges after output recombination.
Formally, a common instantiation uses isolated trajectories $K_t \equiv S$ corresponding to decomposed targets $\{T_1,\dots,T_S\}$, with $\Omega$ aggregating per-session outputs.
TBA distributes exploration across related branches, whereas SSA distributes target components across isolated sessions.
This cross-session fragility is consistent with recent prompt-leakage attacks in multi-tenant LLM deployments, where security failures arise from weak isolation or insufficient global context awareness across interacting components~\cite{wu2025know,hui2024pleak}.

Wahr\'{e}us et al.~\cite{Wahreus2025Prompt} segment prompts into high-level functions and recombine the responses into code, whereas Srivastav et al.~\cite{Srivastav2025Safe} use role-based decomposers, answerers, and combiners.
Context-Fractured Decomposition Attacks~\cite{lin2026context} further show artifact-mediated recombination in tool-using agents.
Across these cases, the common pattern is prompt or role decomposition, isolated processing, and aggregation outside individual sessions.
Harmfulness therefore resides in the orchestration layer, requiring cross-session linkage that preserves partial objectives, intermediate outputs, and their aggregation. Evaluations should test whether a defense identifies the composed objective rather than isolated fragments.

\begin{takeawaybox}
\label{takeaway:ss}
SSA distributes harmful intent across multiple isolated sessions, so reliable detection requires cross-session linkage even when no individual interaction contains sufficient evidence of maliciousness. 
\end{takeawaybox}


\section{Mechanism Analysis}
\label{sec:mechanism}

\begin{table}[!t]
    \centering
    \footnotesize
\caption{ASR of DEA across four target models under three multi-turn interaction strategies.}
    \label{tab:pe_mtj_results}
    \resizebox{\linewidth}{!}{
    \begin{tabular}{lcccc}
        \toprule
        \textbf{Strategy} & \LlamaTHREE & \DeepSeek & \GPTFourO & \LlamaTHREETHREE \\
        \midrule
        Direct Attack & 0.340 & 0.430 & 0.120 & 0.050 \\
        Persistence  & 0.370 & 0.450 & 0.180 & 0.090 \\
        \textbf{FITD} & \textbf{0.960} & \textbf{0.820} & \textbf{0.710} & \textbf{0.440} \\
        \bottomrule
    \end{tabular}
    }
\end{table}

\begin{table}[!t]
    \centering
    \footnotesize
    \setlength{\tabcolsep}{2pt}
    \caption{Final ASR under different malice-level scheduling strategies across target models.}
    \label{tab:strategy-asr}
    \resizebox{\linewidth}{!}{
    \begin{tabular}{cccccccc}
        \toprule
        \multirow{2}{*}{\textbf{Strategy}} & \multicolumn{7}{c}{\textbf{Target Models}} \\
        \cmidrule(lr){2-8}
        & \VicunaSeven & \VicunaThirteen & \MISTRAL & \ChatGLM & \QWENTwo & \LlamaTHREE & \GemmaNine \\
        \midrule
        \textbf{Level Progression} & 0.960 & 1.000 & 1.000 & 0.740 & 1.000 & 0.820 & 0.720 \\
        \textbf{Low-Level Repetition} & 0.960 & 1.000 & 0.900 & 0.660 & 0.960 & 0.460 & 0.480 \\
        \textbf{Random Level Order} & 0.240 & 0.280 & 1.000 & 0.600 & 0.940 & 0.480 & 0.660 \\
        \bottomrule
    \end{tabular}
    }
\end{table}

\subsection{Scope of Analysis}
\label{sec:scope}
\noindent\textbf{Objective.}
This section uses controlled experiments to address the four questions posed in Subsection~\ref{sec:taxonomy}, examining how intent organization, context accumulation, escalation shape, amplification components, and strategy refinement affect attack effectiveness.

\noindent\textbf{Empirical Coverage.}
The experiments cover DEA and SRA, the two single-trajectory categories whose intent-organization mechanisms can be isolated under controlled ablation using available implementations.
Although some TBA and SSA implementations are public, we did not identify implementations suitable for the controlled component-level ablations required here; we therefore systematize these dimensions analytically in Subsection~\ref{sec:taxonomy}.
This asymmetry reflects the limited availability of reproducible, ablation-ready implementations for attacks involving cross-session coordination, where current evaluation has the least coverage.

\noindent\textbf{Competing Hypotheses.}
Two explanations compete for multi-turn attack effectiveness.
The \emph{accumulation} hypothesis attributes success to the volume of context, regardless of how it is organized.
The \emph{organization} hypothesis attributes success to the structure of intent across the trajectory, even when context volume is held fixed.
The analysis proceeds along two axes.
\textbf{RQ1} and~\textbf{RQ3} address what drives effectiveness: \textbf{RQ1} adjudicates between the two hypotheses, and \textbf{RQ3} refines the result by examining how escalation shape interacts with target robustness.
\textbf{RQ2} and~\textbf{RQ4} address what amplifies and scales organized intent, respectively.
The experimental configuration shared across all four analyses is specified in Appendix~\ref{app:setup}.

\begin{table}[!t]
    \centering
    \footnotesize
    \setlength{\tabcolsep}{2pt}
    \caption{Relative ASR gain under different malice-level scheduling strategies across target models.}
    \label{tab:strategy-gain}
    \resizebox{\linewidth}{!}{
    \begin{tabular}{cccccccc}
        \toprule
        \multirow{2}{*}{\textbf{Strategy}} & \multicolumn{7}{c}{\textbf{Target Models}} \\
        \cmidrule(lr){2-8}
        & \VicunaSeven & \VicunaThirteen & \MISTRAL & \ChatGLM & \QWENTwo & \LlamaTHREE & \GemmaNine \\
        \midrule
        \textbf{Level Progression} & 0.900 & 0.840 & 0.320 & 0.740 & 1.000 & 0.820 & 0.700 \\
        \textbf{Low-Level Repetition} & 0.900 & 0.840 & 0.220 & 0.660 & 0.960 & 0.460 & 0.460 \\
        \textbf{Random Level Order} & 0.160 & 0.220 & 0.320 & 0.200 & 0.940 & 0.460 & 0.640 \\
        \bottomrule
    \end{tabular}
    }
\end{table}
 
\begin{table}[!t]
    \centering
    \footnotesize
    \setlength{\tabcolsep}{2pt}
    \caption{Final harmful score under different malice-level scheduling strategies across target models.}
    \label{tab:strategy-harm}
    \resizebox{\linewidth}{!}{
    \begin{tabular}{cccccccc}
        \toprule
        \multirow{2}{*}{\textbf{Strategy}} & \multicolumn{7}{c}{\textbf{Target Models}} \\
        \cmidrule(lr){2-8}
        & \VicunaSeven & \VicunaThirteen & \MISTRAL & \ChatGLM & \QWENTwo & \LlamaTHREE & \GemmaNine \\
        \midrule
        \textbf{Level Progression} & 4.84 & 4.64 & 4.64 & 3.94 & 4.86 & 4.06 & 3.62 \\
        \textbf{Low-Level Repetition} & 4.84 & 4.64 & 3.88 & 3.58 & 4.86 & 3.06 & 2.94 \\
        \textbf{Random Level Order} & 3.26 & 3.48 & 4.64 & 3.26 & 4.08 & 2.86 & 3.14 \\
        \bottomrule
    \end{tabular}
    }
\end{table}

\subsection{Organization versus Accumulation}
\label{sec:org_vs_accum}

Two competing explanations account for multi-turn jailbreak effectiveness.
The \emph{accumulation hypothesis} attributes success to context volume: a longer dialogue history degrades safety behavior regardless of internal organization.
The \emph{organization hypothesis} attributes success to how harmful intent is sequenced across turns, even at constant context volume.
\textbf{RQ1} asks which factor is primary.
We draw evidence from both DEA and SRA, using FITD and MRCJ as representative instantiations.
FITD varies the construction of conversational content across turns, while MRCJ varies the ordering of the attack-strategy sequence; together, they ensure the conclusion does not depend on a single method or a single aspect of the attack process.

\begin{table*}[!t]
    \centering
    \caption{Round-by-round ASR progression for different $\Delta \Theta$ escalation profiles across target models. Colors progress from \protect\colorbox{ProgA}{\strut low} to \protect\colorbox{ProgD}{\strut high}, indicating increasing progression intensity.}
    \label{tab:escalation_results_asr}
    \resizebox{0.93\textwidth}{!}{%
    \renewcommand{\arraystretch}{1.05}%
    \setlength{\tabcolsep}{10pt}%
    \scriptsize
    \begin{tabular}{c|c|c|c|c}
        \toprule
        Model & \textbf{\ChatGLM} & \textbf{\LlamaTHREE} & \textbf{\MISTRAL} & \textbf{\QWENTwo} \\
        \midrule
        $\Delta \Theta$ Sequence
        & \multicolumn{4}{c@{}}{\textbf{ASR sequences: \{$\text{ASR}_{\Theta_1}$, $\text{ASR}_{\Theta_2}$, $\text{ASR}_{\Theta_3}$, $\text{ASR}_{\Theta_4}$\}}} \\
        \midrule
        \{+1, +1, +1, +1\} & \{\asra{0.12}, \asrd{0.98}, \asrd{0.90}, \asrd{0.96}\} & \{\asra{0.00}, \asra{0.02}, \asrc{0.60}, \asrc{0.68}\} & \{\asrd{0.80}, \asrd{1.00}, \asrd{1.00}, \asrd{0.98}\} & \{\asra{0.20}, \asrd{1.00}, \asrc{0.72}, \asrd{1.00}\} \\
        \{+1, +0, +1, +2\} & \{\asra{0.12}, \asrb{0.38}, \asrd{0.94}, \asrd{0.96}\} & \{\asra{0.02}, \asra{0.06}, \asrd{0.94}, \asrd{0.96}\} & \{\asrc{0.54}, \asrc{0.58}, \asrd{0.92}, \asrd{0.92}\} & \{\asra{0.20}, \asrc{0.54}, \asra{0.12}, \asrb{0.28}\} \\
        \{+1, +2, +1, +0\} & \{\asrb{0.42}, \asrd{0.80}, \asrd{0.86}, \asrc{0.62}\} & \{\asra{0.02}, \asrb{0.42}, \asrc{0.68}, \asrd{0.86}\} & \{\asrd{0.80}, \asrd{1.00}, \asrd{1.00}, \asrd{1.00}\} & \{\asra{0.16}, \asrc{0.64}, \asrc{0.72}, \asrd{0.84}\} \\
        \{+1, +0, +2, +1\} & \{\asra{0.12}, \asrb{0.40}, \asrd{0.88}, \asrd{0.98}\} & \{\asra{0.02}, \asra{0.06}, \asrb{0.44}, \asra{0.20}\} & \{\asrb{0.38}, \asrc{0.60}, \asrc{0.70}, \asrc{0.66}\} & \{\asrb{0.28}, \asrb{0.30}, \asra{0.22}, \asrc{0.74}\} \\
        \{+1, +1, +0, +2\} & \{\asra{0.12}, \asrd{0.98}, \asrd{0.96}, \asrd{0.96}\} & \{\asra{0.02}, \asra{0.04}, \asrd{0.82}, \asrd{0.88}\} & \{\asrb{0.38}, \asrd{1.00}, \asrd{0.98}, \asrd{1.00}\} & \{\asrb{0.36}, \asrd{0.90}, \asra{0.18}, \asra{0.10}\} \\
        \{+1, +1, +2, +0\} & \{\asra{0.10}, \asrd{1.00}, \asrd{0.90}, \asrd{1.00}\} & \{\asra{0.02}, \asra{0.14}, \asrc{0.62}, \asrc{0.54}\} & \{\asrd{0.80}, \asrd{1.00}, \asrd{0.98}, \asrd{1.00}\} & \{\asra{0.20}, \asrd{0.90}, \asrd{1.00}, \asrd{1.00}\} \\
        \{+1, +2, +0, +1\} & \{\asra{0.12}, \asrd{0.92}, \asrd{0.98}, \asrd{0.90}\} & \{\asra{0.02}, \asra{0.10}, \asrc{0.74}, \asrd{0.86}\} & \{\asrb{0.38}, \asrd{0.96}, \asrd{0.90}, \asrd{0.92}\} & \{\asra{0.20}, \asrc{0.60}, \asrc{0.72}, \asrc{0.72}\} \\
        \midrule
        Model & \textbf{\VicunaSeven} & \textbf{\VicunaThirteen} & \textbf{\GemmaNine} & \textbf{Average} \\
        \midrule
        $\Delta \Theta$ Sequence
        & \multicolumn{4}{c@{}}{\textbf{ASR sequences: \{$\text{ASR}_{\Theta_1}$, $\text{ASR}_{\Theta_2}$, $\text{ASR}_{\Theta_3}$, $\text{ASR}_{\Theta_4}$\}}} \\
        \midrule
        \{+1, +1, +1, +1\} & \{\asrd{0.98}, \asrd{1.00}, \asrd{1.00}, \asrd{1.00}\} & \{\asrd{0.84}, \asrd{1.00}, \asrd{0.98}, \asrd{0.98}\} & \{\asra{0.08}, \asrb{0.36}, \asrc{0.58}, \asrc{0.72}\} & \{\asrb{0.43}, \asrd{0.77}, \asrd{0.83}, \asrd{0.90}\} \\
        \{+1, +0, +1, +2\} & \{\asrd{0.98}, \asrd{0.98}, \asrd{1.00}, \asrd{1.00}\} & \{\asrd{0.84}, \asrd{0.94}, \asrd{1.00}, \asrd{0.94}\} & \{\asra{0.04}, \asra{0.10}, \asrc{0.62}, \asrd{0.80}\} & \{\asrb{0.39}, \asrc{0.51}, \asrd{0.79}, \asrd{0.84}\} \\
        \{+1, +2, +1, +0\} & \{\asrd{0.98}, \asrd{1.00}, \asrd{0.98}, \asrd{0.90}\} & \{\asrd{0.84}, \asrd{0.94}, \asrd{0.92}, \asrd{0.82}\} & \{\asra{0.12}, \asrb{0.38}, \asrc{0.50}, \asrc{0.56}\} & \{\asrb{0.48}, \asrc{0.74}, \asrd{0.81}, \asrd{0.80}\} \\
        \{+1, +0, +2, +1\} & \{\asrd{0.98}, \asrd{0.98}, \asrd{1.00}, \asrd{0.98}\} & \{\asrd{0.84}, \asrd{0.94}, \asrd{1.00}, \asrd{1.00}\} & \{\asra{0.04}, \asra{0.08}, \asrb{0.32}, \asrb{0.38}\} & \{\asrb{0.38}, \asrb{0.48}, \asrc{0.65}, \asrc{0.71}\} \\
        \{+1, +1, +0, +2\} & \{\asrd{0.98}, \asrd{1.00}, \asrd{0.94}, \asrd{1.00}\} & \{\asrd{0.84}, \asrd{1.00}, \asrd{1.00}, \asrd{0.86}\} & \{\asra{0.06}, \asra{0.10}, \asrb{0.42}, \asrc{0.74}\} & \{\asrb{0.39}, \asrc{0.72}, \asrd{0.76}, \asrd{0.79}\} \\
        \{+1, +1, +2, +0\} & \{\asrd{1.00}, \asrd{1.00}, \asrd{1.00}, \asrd{0.86}\} & \{\asrd{0.84}, \asrd{1.00}, \asrd{0.98}, \asrd{0.96}\} & \{\asra{0.08}, \asra{0.18}, \asrb{0.44}, \asrc{0.50}\} & \{\asrb{0.44}, \asrd{0.75}, \asrd{0.85}, \asrd{0.84}\} \\
        \{+1, +2, +0, +1\} & \{\asrd{0.98}, \asrd{1.00}, \asrd{0.90}, \asrd{0.90}\} & \{\asrd{0.94}, \asrd{0.96}, \asrd{1.00}, \asrd{1.00}\} & \{\asra{0.06}, \asra{0.16}, \asrb{0.48}, \asrc{0.64}\} & \{\asrb{0.39}, \asrc{0.67}, \asrd{0.82}, \asrd{0.85}\} \\
        \bottomrule
    \end{tabular}
    }
\end{table*}

\noindent\textbf{DEA-side evidence: FITD.}
\label{sec:experiments_pe_mtj}
We evaluate FITD-style DEA~\cite{weng2025foot} on four target models (\LlamaTHREE, \DeepSeek, \GPTFourO, and \LlamaTHREETHREE).
For each model we compare three interaction strategies that share the same target queries and differ only in how the query sequence is organized:
\begin{itemize}
    \item \textbf{Direct Attack.} The malicious query is presented in a zero-shot fashion.
    \item \textbf{Persistence Strategy.} Upon refusal, the attacker retries the same request with minor rephrasing for a fixed number of attempts, without introducing benign context.
    \item \textbf{FITD.} The conversation starts from safe, related topics and progressively increases sensitivity before posing the target query.
\end{itemize}
The Persistence condition produces a multi-turn interaction of comparable length to the Escalation condition but without organizing its content toward the harmful objective, thereby isolating the effect of content organization from that of interaction length.

As \autoref{tab:pe_mtj_results} shows, Direct Attack and Persistence yield comparable Attack Success Rate (ASR) across all four models, while FITD-style escalation achieves consistently higher rates regardless of alignment strength.
This pattern suggests that organizing conversational content across turns provides a benefit beyond simply issuing more queries.

\noindent\textbf{SRA-side evidence: MRCJ.}
\label{sec:semtj_evidence_block}
We draw complementary evidence from the MRCJ methodology, which controls a different aspect of the attack: the ordering of the strategy sequence rather than the content of individual turns.
We evaluate seven safety-aligned target models using auxiliary questions from \textbf{MUCD}, which labels each question with a discrete malice level, and final target questions from \textbf{AdvBench}~\cite{zou2023universal}.

We compare three malice-level schedules for the auxiliary-question sequence, holding turn count and available malice levels fixed:
\begin{itemize}
    \item \textbf{Level Progression.} Randomly sampled auxiliary questions are arranged by strictly increasing malice level.
    \item \textbf{Low-Level Repetition.} Low-malice auxiliary turns are reused throughout the interaction, producing the same context volume without escalation.
    \item \textbf{Random Level Order.} Randomly sampled auxiliary questions are arranged across the available malice levels in a non-monotonic order, preserving the level inventory without an escalation schedule.
\end{itemize}
As \autoref{tab:strategy-asr}, \autoref{tab:strategy-gain}, and \autoref{tab:strategy-harm} show, \emph{Level Progression} achieves the strongest ASR, relative ASR gain, and harmfulness scores across all models, with the gap most pronounced on robustly aligned targets such as \ChatGLM and \LlamaTHREE.
Because \emph{Low-Level Repetition} provides a matched interaction length and \emph{Random Level Order} preserves the malice-level inventory, these results support the role of the escalation schedule beyond the matched alternatives.

\noindent\textbf{Cross-dimension synthesis.}
The two experiments examine different intent-organization dimensions yet converge on the same conclusion.
FITD shows that organized escalation of conversational content outperforms request repetition at comparable interaction length; MRCJ shows that Level Progression outperforms both Low-Level Repetition and Random Level Order at the same turn count and malice-level inventory.
Across these tested implementations, organized sequencing is associated with higher effectiveness than the matched alternatives that primarily increase or reorder accumulated context. This pattern appears in both content-level query organization and strategy-level sequence organization.

\begin{findingbox}
In the tested FITD and MRCJ settings, the observed advantage is consistent with a contribution from intent organization beyond accumulated context volume.
\end{findingbox}

\subsection{Escalation Shape across Target Models}
\label{sec:escalation_shape}

Building on the RQ1 results in Subsection~\ref{sec:org_vs_accum}, we next examine the form that organized escalation takes.
Organized escalation is not monolithic, however.
The strategy sequence can follow different increment profiles, and \textbf{RQ3} asks whether a single profile universally suffices or the optimal shape depends on the target model.

Using the same MRCJ framework and models as Subsection~\ref{sec:org_vs_accum}, we fix the endpoint malice level and the number of escalation steps at four, varying only how the total increment is distributed across steps.
All conditions are monotonically non-decreasing and differ only in escalation shape, providing a finer-grained comparison than the organized-versus-unorganized contrast of Subsection~\ref{sec:org_vs_accum}.

\begin{table*}[!t]
    \centering
    \caption{Round-by-round harmfulness score progression for different $\Delta \Theta$ escalation profiles across target models. Colors progress from \protect\colorbox{ProgA}{\strut low} to \protect\colorbox{ProgD}{\strut high}, indicating increasing progression intensity.}
    \label{tab:escalation_results_score}
    \resizebox{0.92\textwidth}{!}{%
    \renewcommand{\arraystretch}{1.05}%
    \setlength{\tabcolsep}{10pt}%
    \scriptsize
    \begin{tabular}{c|c|c|c|c}
        \toprule
        
        Model & \textbf{\ChatGLM} & \textbf{\LlamaTHREE} & \textbf{\MISTRAL} & \textbf{\QWENTwo} \\
        \midrule
        $\Delta \Theta$ Sequence
        & \multicolumn{4}{c}{\textbf{Harmful score sequences: \{$\text{Score}_{\Theta_1}$, $\text{Score}_{\Theta_2}$, $\text{Score}_{\Theta_3}$, $\text{Score}_{\Theta_4}$\}}} \\
        \midrule
        \{+1, +1, +1, +1\} & \{\scorec{3.08}, \scorec{3.22}, \scorec{3.86}, \scorec{3.94}\} & \{\scorea{1.00}, \scorea{1.08}, \scorec{3.02}, \scorec{3.26}\} & \{\scored{4.14}, \scored{4.64}, \scored{4.82}, \scored{4.76}\} & \{\scorea{1.80}, \scored{4.48}, \scorec{3.64}, \scored{4.04}\} \\
        \{+1, +0, +1, +2\} & \{\scorea{1.40}, \scoreb{2.50}, \scored{4.58}, \scored{4.46}\} & \{\scorea{1.06}, \scorea{1.20}, \scored{4.36}, \scored{4.54}\} & \{\scorec{3.04}, \scorec{3.10}, \scored{4.14}, \scored{4.00}\} & \{\scorea{1.78}, \scorec{3.38}, \scoreb{2.46}, \scoreb{2.76}\} \\
        \{+1, +2, +1, +0\} & \{\scoreb{2.46}, \scorec{3.58}, \scorec{3.94}, \scorec{3.16}\} & \{\scorea{1.08}, \scoreb{2.48}, \scorec{3.26}, \scorec{3.74}\} & \{\scored{4.12}, \scored{4.78}, \scored{4.82}, \scored{4.82}\} & \{\scorea{1.68}, \scorec{3.44}, \scorec{3.62}, \scorec{3.84}\} \\
        \{+1, +0, +2, +1\} & \{\scorea{1.40}, \scoreb{2.52}, \scored{4.30}, \scored{4.48}\} & \{\scorea{1.08}, \scorea{1.22}, \scoreb{2.74}, \scorea{1.78}\} & \{\scoreb{2.50}, \scorec{3.14}, \scorec{3.70}, \scorec{3.66}\} & \{\scoreb{2.00}, \scorec{3.06}, \scoreb{2.46}, \scorec{3.58}\} \\
        \{+1, +1, +0, +2\} & \{\scorea{1.40}, \scored{4.10}, \scored{4.14}, \scored{4.02}\} & \{\scorea{1.08}, \scorea{1.16}, \scorec{3.76}, \scored{4.12}\} & \{\scoreb{2.50}, \scored{4.96}, \scored{4.48}, \scored{4.62}\} & \{\scoreb{2.24}, \scorec{3.96}, \scorec{3.10}, \scorec{3.10}\} \\
        \{+1, +1, +2, +0\} & \{\scorea{1.38}, \scored{4.38}, \scorec{3.92}, \scored{4.20}\} & \{\scorea{1.08}, \scorea{1.42}, \scorec{3.20}, \scoreb{2.88}\} & \{\scored{4.10}, \scored{4.62}, \scored{4.38}, \scored{4.26}\} & \{\scorea{1.82}, \scored{4.40}, \scored{4.52}, \scored{4.02}\} \\
        \{+1, +2, +0, +1\} & \{\scorea{1.40}, \scored{4.02}, \scored{4.26}, \scored{4.06}\} & \{\scorea{1.08}, \scorea{1.36}, \scorec{3.88}, \scored{4.26}\} & \{\scoreb{2.52}, \scored{4.26}, \scored{4.06}, \scored{4.12}\} & \{\scorea{1.82}, \scorec{3.56}, \scorec{3.70}, \scorec{3.66}\} \\
        \midrule
        
        Model & \textbf{\VicunaSeven} & \textbf{\VicunaThirteen} & \textbf{\GemmaNine} & \textbf{Average} \\
        \midrule
        $\Delta \Theta$ Sequence
        & \multicolumn{4}{c@{}}{\textbf{Harmful score sequences: \{$\text{Score}_{\Theta_1}$, $\text{Score}_{\Theta_2}$, $\text{Score}_{\Theta_3}$, $\text{Score}_{\Theta_4}$\}}} \\
        \midrule
        \{+1, +1, +1, +1\} & \{\scored{4.88}, \scored{4.94}, \scored{4.98}, \scored{4.98}\} & \{\scored{4.14}, \scored{4.92}, \scored{4.72}, \scored{4.76}\} & \{\scorea{1.28}, \scoreb{2.42}, \scorec{3.18}, \scorec{3.62}\} & \{\scoreb{2.90}, \scorec{3.68}, \scored{4.03}, \scored{4.19}\} \\
        \{+1, +0, +1, +2\} & \{\scored{4.88}, \scored{4.80}, \scored{4.96}, \scored{4.96}\} & \{\scored{4.14}, \scored{4.68}, \scored{4.82}, \scored{4.68}\} & \{\scorea{1.12}, \scorea{1.36}, \scorec{3.34}, \scorec{3.86}\} & \{\scoreb{2.49}, \scorec{3.00}, \scored{4.09}, \scored{4.18}\} \\
        \{+1, +2, +1, +0\} & \{\scored{4.88}, \scored{4.98}, \scored{4.90}, \scored{4.52}\} & \{\scored{4.24}, \scored{4.66}, \scored{4.64}, \scored{4.20}\} & \{\scorea{1.42}, \scoreb{2.50}, \scoreb{2.88}, \scorec{3.06}\} & \{\scoreb{2.84}, \scorec{3.77}, \scored{4.01}, \scorec{3.91}\} \\
        \{+1, +0, +2, +1\} & \{\scored{4.88}, \scored{4.78}, \scored{4.92}, \scored{4.86}\} & \{\scored{4.28}, \scored{4.54}, \scored{4.72}, \scored{4.80}\} & \{\scorea{1.14}, \scorea{1.26}, \scoreb{2.28}, \scoreb{2.46}\} & \{\scoreb{2.47}, \scoreb{2.93}, \scorec{3.59}, \scorec{3.66}\} \\
        \{+1, +1, +0, +2\} & \{\scored{4.88}, \scored{4.94}, \scored{4.70}, \scored{4.94}\} & \{\scored{4.12}, \scored{4.90}, \scored{4.82}, \scored{4.38}\} & \{\scorea{1.20}, \scorea{1.38}, \scoreb{2.68}, \scorec{3.70}\} & \{\scoreb{2.50}, \scorec{3.63}, \scorec{3.95}, \scored{4.13}\} \\
        \{+1, +1, +2, +0\} & \{\scored{4.94}, \scored{4.92}, \scored{4.98}, \scored{4.36}\} & \{\scored{4.14}, \scored{4.92}, \scored{4.86}, \scored{4.78}\} & \{\scorea{1.26}, \scorea{1.62}, \scoreb{2.72}, \scoreb{2.92}\} & \{\scoreb{2.67}, \scorec{3.75}, \scored{4.08}, \scorec{3.92}\} \\
        \{+1, +2, +0, +1\} & \{\scored{4.88}, \scored{4.98}, \scored{4.56}, \scored{4.58}\} & \{\scored{4.68}, \scored{4.74}, \scored{4.98}, \scored{4.98}\} & \{\scorea{1.20}, \scorea{1.58}, \scoreb{2.82}, \scorec{3.36}\} & \{\scoreb{2.51}, \scorec{3.51}, \scored{4.04}, \scored{4.15}\} \\
        \bottomrule
    \end{tabular}
    }
\end{table*}

\begin{table}[!h]
    \centering
    \caption{Efficiency analysis of malice-increment profiles: query times (QT) required by each schedule (lower is better), where one user--model question--answer exchange counts as one QT. The shaded cells represent the minimum QT achieved for each model.}
    \label{tab:escalation_results_qt}
    \footnotesize
    \resizebox{\linewidth}{!}{%
    \renewcommand{\arraystretch}{1.05}%
    \setlength{\tabcolsep}{1.5pt}%
    \footnotesize
    \begin{tabular}{lccccccc}
        \toprule
        \textbf{$\Delta \Theta$ Sequence} & \ChatGLM & \LlamaTHREE & \MISTRAL & \QWENTwo & \VicunaSeven & \VicunaThirteen & \GemmaNine \\
        \midrule
        \{+1, +1, +1, +1\} & 87 & 262 & \highlightcell{57} & 147 & 57 & 56 & 156 \\
        \{+1, +0, +1, +2\} & 108 & \highlightcell{130} & 63 & \highlightcell{119} & 57 & 61 & \highlightcell{128} \\
        \{+1, +2, +1, +0\} & 169 & 304 & 56 & 224 & 56 & 56 & 231 \\
        \{+1, +0, +2, +1\} & 112 & 625 & 83 & 351 & 59 & 62 & 212 \\
        \{+1, +1, +0, +2\} & 86 & 180 & 82 & 174 & 59 & \highlightcell{55} & 144 \\
        \{+1, +1, +2, +0\} & \highlightcell{84} & 289 & \highlightcell{57} & 124 & 57 & 56 & 168 \\
        \{+1, +2, +0, +1\} & 96 & 171 & 64 & 284 & \highlightcell{56} & \highlightcell{55} & 192 \\
        \bottomrule
    \end{tabular}
    }
\end{table}

\noindent\textbf{The optimal profile varies across target models.}
\autoref{tab:escalation_results_asr} shows that on average across all targets, the uniform schedule $\{+1,+1,+1,+1\}$ achieves the strongest final ASR, indicating that evenly paced escalation is a robust default when target-specific robustness is not known.
For \VicunaSeven, \VicunaThirteen, and \MISTRAL, most increment profiles reach high final ASR; the uniform schedule $\{+1,+1,+1,+1\}$ already suffices.
For \ChatGLM, \LlamaTHREE, and \GemmaNine, profile shape is more consequential: schedules with a stabilization phase followed by a concentrated late jump, such as $\{+1,0,+1,+2\}$, can outperform uniform escalation.
As shown in~\autoref{tab:escalation_results_score}, the round-by-round harmful scores suggest a mechanism: effective non-uniform profiles maintain low harmful scores in early rounds, reducing the risk of premature defense activation, and concentrate the escalation in the final step.

\noindent\textbf{Effective profiles tend to be efficient.}
For some target models, \autoref{tab:escalation_results_qt} shows that the uniform schedule also uses the fewest queries.
For others, non-uniform profiles that yield higher ASR can also use fewer queries, indicating that matching escalation shape to the target model can improve both effectiveness and efficiency.

\begin{findingbox}
Uniform escalation is a strong aggregate default, but effective intent organization may still require adapting the escalation shape to the target model.
\end{findingbox}

\subsection{Amplification within Organized Intent}
\label{sec:amplification}

\textbf{RQ2} asks to what extent amplification components embedded within organized DEA trajectories independently contribute to effectiveness.
Methods in this category commonly combine components such as diversified semantic pathways and multi-step reasoning chains, yet existing evaluations introduce them together without isolating individual contributions.
ActorAttack~\cite{ren2024derail} is considered a representative instantiation in this paper because its design cleanly separates the two components, enabling independent ablation.
We evaluate on three target models (\DeepSeek, \GPTFourO, and \LlamaTHREETHREE).

\begin{figure}[t]
    \centering
    \includegraphics[width=\linewidth]{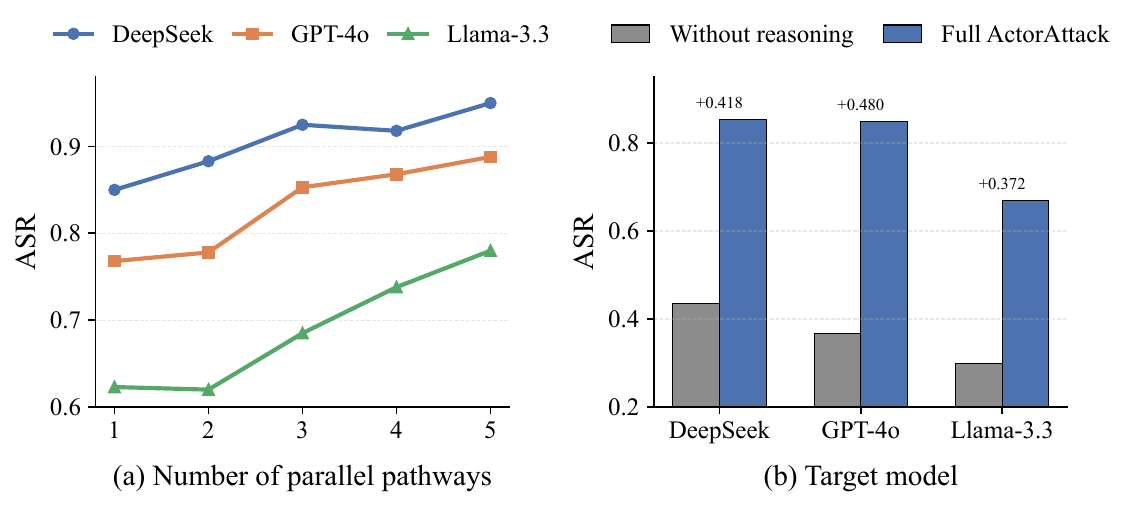}
    \caption{
    Ablation study of two key aspects of ActorAttack.
    (a) Increasing the number of parallel semantic pathways consistently improves attack success across target models.
    (b) Removing reasoning chains substantially reduces attack success, indicating that progressive reasoning contributes more to attack effectiveness than semantic diversity alone.
    }
    \label{fig:actorattack_ablation}
\end{figure}

\noindent\textbf{Semantic diversity and reasoning chains play complementary roles.}
ActorAttack combines two orthogonal mechanisms for organizing harmful intent: parallel semantic pathways broaden the search over diverse conversational framings, while reasoning chains progressively contextualize harmful intent within each pathway.
To quantify the contributions, we separately vary the number of semantic pathways and remove reasoning chains while holding all remaining components fixed, as demonstrated in \autoref{fig:actorattack_ablation}.

\autoref{fig:actorattack_ablation}a shows that increasing semantic pathway diversity consistently improves attack success across all evaluated models, indicating that different pathways expose complementary vulnerabilities by presenting the same malicious objective through semantically distinct conversational trajectories.
The improvement, however, gradually saturates, suggesting that additional pathways mainly increase the probability of discovering an effective attack trajectory rather than fundamentally changing how harmful intent is organized.
By contrast, \autoref{fig:actorattack_ablation}b shows that removing reasoning chains leads to a substantial degradation in attack success across all target models.
This observation indicates that reasoning chains constitute the primary mechanism underlying ActorAttack: rather than exposing harmful intent directly, they progressively establish a legitimate semantic context in which the final malicious objective appears as the natural conclusion of an otherwise benign reasoning process.
Together, these results suggest that semantic diversity expands the attack surface, whereas reasoning chains provide a stronger mechanism for progressively embedding harmful intent.

\noindent\textbf{Components contribute independently through complementary mechanisms.}
Semantic diversity explores alternative framings of the harmful objective, while reasoning chains increase the persuasiveness of each framing.
Within ActorAttack, reasoning chains account for the larger share, though the relative balance may differ across methods.

\begin{findingbox}
Amplification components within organized trajectories further enhance attack effectiveness through complementary mechanisms. Multi-turn attack effectiveness is therefore compositional: trajectory-level intent organization provides the foundation, while method-specific components amplify its effect.
\end{findingbox}

\subsection{Sub-Paradigm Progression}
\label{sec:subparadigm_progression}

\textbf{RQ4} asks whether the three SRA sub-paradigms, Fixed-strategy, Adaptive-optimization, and Agent-based, correspond to genuinely distinct capability levels or are interchangeable implementations at comparable effectiveness.

\noindent\textbf{Same-backbone sub-paradigm chain.}
To compare the SRA sub-paradigms within a common implementation context, we construct a progressive configuration chain on a single backbone, X-Teaming~\cite{rahman2025x}.
The detailed construction is reported in Appendix~\ref{app:rq4} and \autoref{tab:xt-components}. A1, A2, and A3 instantiate the Fixed-strategy, Adaptive-optimization, and Agent-based sub-paradigms, respectively. Moving along the chain jointly changes the planning resources and decision mechanisms characteristic of each sub-paradigm; it is therefore a comparison of sub-paradigm configurations rather than an isolation of individual components.
We evaluate on \DeepSeek, \GPTFourO, \LlamaTHREETHREE, and \GeminiThree; all conditions share a fixed turn budget and constant retry count, avoiding differences in interaction budget.

\begin{table}[t]
    \centering
    \caption{ASR on the main chain across four target models. Samples are drawn from JailbreakBench.}
    \label{tab:xt-mainchain}
    \footnotesize
    \setlength{\tabcolsep}{6pt}
    \renewcommand{\arraystretch}{1.1}
    \resizebox{0.95\linewidth}{!}{%
    \begin{tabular}{@{}ccccc@{}}
        \toprule
        \textbf{Setting} & \textbf{\DeepSeek} & \textbf{\GPTFourO} & \textbf{\LlamaTHREETHREE} & \textbf{\GeminiThree}\\
        \midrule
        \textbf{A1} & 0.770 & 0.540 & 0.470 & 0.270\\
        \textbf{A2} & 0.810 & 0.640 & 0.600 & 0.340\\
        \textbf{A3} & 0.980 & 0.820 & 0.700 & 0.470\\
        \bottomrule
    \end{tabular}%
    }
\end{table}

\noindent\textbf{The three sub-paradigms form a genuine hierarchy.}
As \autoref{tab:xt-mainchain} shows, ASR increases monotonically from A1 to A3 on every target, confirming that the three sub-paradigms correspond to distinct capability levels rather than interchangeable variants.
The magnitude of the gain varies across target models, indicating that the value of stronger inference-time decision-making is target-dependent.

\begin{findingbox}
The three SRA sub-paradigms exhibit a consistent capability hierarchy in this ablation.
The magnitude of the gains depends on the target model.
\end{findingbox}

\section{Discussion}
\label{sec:discussion}

\noindent\textbf{Intent Analysis and Observation Scope.}
Our taxonomy classifies multi-turn jailbreaks by where \emph{harmful intent} is organized, and the mechanism analysis shows that this organization determines the required \emph{observation scope}.
DEA and SRA commonly require session-level observation of dialogue history and, for SRA, evolving strategy state; TBA requires joint analysis of related branches, while SSA requires cross-session linkage and aggregation.
For hybrid methods, the required observation scope is determined by the composition of their active \emph{intent-organization dimensions} rather than by a single category label.
This compositional view highlights the limitations of turn-local defenses and single-point ASR evaluation: monitoring should retain links among observations and report both its observation scope and the earliest point at which the organized harmful objective is identifiable.
A broader scope also need not delay intervention until a dialogue is complete. A monitor can update a risk assessment incrementally as linked turns arrive and escalate review once accumulated evidence crosses a deployment-defined threshold.
Such incremental monitoring should be evaluated not only for eventual detection accuracy but also for the timeliness and stability of its intervention decisions as additional context becomes available.

\noindent\textbf{Multimodal and Tool-Mediated Extensions.}
The intent-organization dimensions abstract over input modality and execution environment: they characterize how \emph{harmful intent} is organized across interactions, whereas multimodality and tool mediation characterize the channels and operational settings through which an attack is realized.
While multimodal and tool-mediated jailbreaks are actively studied more broadly~\cite{liang2026t2vshield,liu2025t2v}, only a small and heterogeneous subset of the multi-turn corpus exhibits these attributes.
Recent multi-turn examples include target-side multimodality and multimodal prompting~\cite{das2026multi, choi2026multi, zhao2026when}, whereas STAC and CFD illustrate attacks mediated by tool chains or persistent artifacts~\cite{li2025stac, lin2026context}.
Treating either attribute as an additional top-level dimension would therefore conflate \emph{operational conditions} with \emph{intent organization} and overstate the available multi-turn evidence.
Future work should examine how harmful intent persists across text, images, tool outputs, and artifacts, and should develop evaluations that link cross-modal context with tool and artifact provenance over extended interactions.
Such evaluations should distinguish \emph{a harmful intent signal} from \emph{the channel carrying it}, so that changing modality or delegating a step to a tool does not obscure responsibility for the overall trajectory.
They should also test whether safety controls preserve this provenance when intermediate outputs are transformed, summarized, or stored for later reuse.

\noindent\textbf{Future Directions.} 
Future work should develop interaction-aware benchmarks that retain multi-label mechanism annotations and evaluate the \emph{observation scopes} at which intent becomes detectable, extending recent efforts such as SafeDialBench and AgentHarm~\cite{cao2026safedialbench,andriushchenko2025agentharm}. 
As LLM systems acquire persistent memory, tools, and autonomy, defenses should combine monitoring across content, strategy, branches, and sessions~\cite{shahroz2025agents,zhang2025agent}. 
Standardized implementations and datasets for \emph{trajectory-branching and session-splitting attacks} remain particularly needed~\cite{song2026multibreak}.

\section{Conclusion}
Multi-turn jailbreaks elevate the threat from single-prompt vulnerabilities to long-horizon control over dialogue trajectories. 
We present an interaction-centric taxonomy organized around four non-exclusive intent-organization dimensions and targeted empirical analyses of representative mechanisms.
The framework explains how distinct attacks can share interaction patterns while differing in their control points for organizing harmful intent.
Across the tested single-trajectory settings, deliberately organized intent outperforms matched alternatives that primarily increase accumulated context.
It motivates evaluation and defense mechanisms grounded in the combined observation requirements of active dimensions.
We hope this taxonomy provides a foundation for systematic comparison across methods and motivates future defenses for advanced attacks against large foundation models.

{\small
\bibliographystyle{IEEEtran}
\bibliography{ref}
}

\appendices

\begin{table*}[t]
\centering
\tiny
\setlength{\tabcolsep}{2.4pt}
\renewcommand{\arraystretch}{0.82}
\caption[Method-level taxonomy of multi-turn jailbreak methods.]{Method-level taxonomy of multi-turn jailbreak methods.
\textbf{Type} records non-exclusive labels; methods receive multiple labels when their attacks instantiate multiple dimensions.
\textbf{Interaction Scope} distinguishes one trajectory from multiple trajectories or sessions. \textbf{Observation Scope} states the minimum context required to expose harmful intent: \textbf{Turn}, \textbf{Session}, or \textbf{Cross-session}, the latter including linked sessions and related branches.
\textbf{Auxiliary Capability} records supporting resources: \textbf{Tool} includes retrieval, optimization, code aggregation, APIs, and memory/artifacts; \textbf{Multimodal} refers to non-textual inputs or targets; \textbf{None} indicates no additional capability.
\textbf{Code}: \full{} indicates open-source code; \none{} indicates no public implementation.}
\label{tab:method_level_taxonomy}
\resizebox{\textwidth}{!}{%
\begin{tabular}{@{}ccccccc@{}}
\toprule
\textbf{Method} & \textbf{Type} & \textbf{Attack Mechanism} & \textbf{Interaction Scope} & \textbf{Observation Scope} & \textbf{Auxiliary Capability} & \textbf{Code} \\
\midrule
Crescendo~\cite{russinovich2025great} & DEA+SRA & LLM & Single & Session & None & \full \\
FITD~\cite{weng2025foot} & DEA & Template/LLM & Single & Session & None & \full \\
CFA~\cite{Sun2024} & DEA & Template & Single & Session & None & \none \\
AMA~\cite{wuanalogy} & DEA & Template & Single & Session & None & \full \\
CIA~\cite{Cheng2024} & DEA & LLM & Single & Session & None & \none \\
ActorAttack~\cite{ren2024derail} & DEA+SRA & LLM & Single & Session & None & \full \\
RACE~\cite{Ying2025} & DEA+SRA & LLM & Single & Session & None & \none \\
ICON~\cite{lin2026icon} & DEA+SRA & LLM & Single & Session & None & \full \\
Imposter.AI~\cite{Liu2024} & DEA & Template & Single & Session & None & \none \\
Jigsaw Puzzles~\cite{yang2025jigsaw} & DEA & Template & Single & Session & None & \full \\
Speak-Out-of-Turn~\cite{Zhou2024} & DEA & LLM & Single & Session & None & \none \\
MRJ-Agent~\cite{Wang2024} & DEA+SRA & LLM agent & Single & Session & None & \none \\
SIREN~\cite{Zhao2025} & DEA+SRA & Fine-tuned LLM & Single & Session & None & \full \\
MRCJ~\cite{MRCJ} & DEA+SRA & Template & Single & Turn & None & \full \\
EMRA~\cite{li2026coopguard} & DEA+SRA & Template & Single & Turn & None & \none \\
ASJA~\cite{du2025multi} & DEA+SRA & LLM/optimizer & Single & Turn & None & \none \\
SoC-MAB~\cite{ramesh2025efficient} & DEA+SRA & RL/judge & Single & Turn & None & \none \\
iMIST~\cite{wang2026jailbreaking} & DEA+SRA & RL policy & Single & Turn & Tool & \none \\
SEMA~\cite{feng2026sema} & DEA+SRA & RL-trained LLM & Single & Turn & None & \full \\
GOAT~\cite{pavlova2024automated} & DEA+SRA & LLM agent & Single & Session & None & \full \\
GALA~\cite{chen2025strategize} & DEA+SRA & LLM agent & Single & Session & None & \full \\
RedCoder~\cite{mo2025redcoder} & DEA+SRA & LLM+RAG & Single & Session & Tool & \full \\
X-Teaming~\cite{rahman2025x} & DEA+SRA & Multi-agents & Single & Session & None & \full \\
Tempest~\cite{zhou2025tempest} & SRA+TBA & LLM/tree search & Multi & Cross-session & Tool & \none \\
SLIP~\cite{kulshreshtha2026multi} & DEA+SRA+TBA & Template/tree search & Multi & Cross-session & Tool & \full \\
MHJ~\cite{li2024llm} & DEA+SRA+TBA & Human & Multi & Cross-session & None & \full \\
ADD~\cite{Srivastav2025Safe} & DEA+SSA & Multi-agents & Multi & Cross-session & None & \full \\
PDC~\cite{Wahreus2025Prompt} & DEA+SRA+SSA & Multi-agents & Multi & Cross-session & Tool & \none \\
Chain of Attack~\cite{yang2025chain} & DEA+SRA & LLM & Single & Session & None & \full \\
Echo Chamber~\cite{alobaid2026echo} & DEA & Template/LLM & Single & Session & None & \full \\
NEXUS~\cite{rafieiasl2025nexus} & DEA+SRA+TBA & Multi-agent graph & Multi & Cross-session & None & \full \\
HarmNet~\cite{narula2025harmnet} & DEA+SRA+TBA & Multi-agent graph & Multi & Cross-session & None & \none \\
PE-CoA~\cite{nihal2026pattern} & DEA & Patterned LLM & Single & Session & None & \full \\
TROJail~\cite{xiong2026trojail} & DEA+SRA & RL-trained LLM & Single & Session & None & \full \\
Mastermind~\cite{li2026knowledge} & DEA+SRA & LLM+knowledge base & Single & Session & Tool & \none \\
JAIL~\cite{feng2026jail} & SRA+TBA & LLM/optimizer & Multi & Cross-session & Tool & \none \\
MJAD-MLLM~\cite{das2026multi} & DEA & Template/LLM & Single & Session & Multimodal & \none \\
MAPA~\cite{choi2026multi} & DEA+SRA & LLM/optimizer & Single & Session & Multimodal & \none \\
STAC~\cite{li2025stac} & DEA+SRA & Tool-chain planner & Single & Session & Tool & \full \\
CFD~\cite{lin2026context} & SSA & Tool pipeline & Multi & Cross-session & Tool & \full \\
Inception~\cite{zhao2026when} & DEA & Segmentation/recursion & Single & Session & Tool + Multimodal & \full \\
GRAF~\cite{tang2025graf} & DEA+SRA & LLM/fabrication & Single & Session & None & \full \\
PLAGUE~\cite{bhuiya2026plague} & DEA+SRA & LLM agent & Single & Session & None & \none \\
Jailbreak Forests~\cite{anonymous2026multiturn} & SRA+TBA & Reasoning agent/RL & Multi & Cross-session & None & \none \\
ABC path search~\cite{liu2025let} & SRA+TBA & Path optimizer & Multi & Cross-session & Tool & \none \\
\bottomrule
\end{tabular}%
}
\end{table*}

\section{Experimental Setup}
\label{app:setup}

This appendix records the execution details for the controlled DEA and SRA analyses in Section~\ref{sec:mechanism}; \autoref{tab:app-exp-map} maps research questions to experimental settings.

\begin{table}[t]
    \centering
    \caption{Mapping between research questions and experimental settings.}
    \label{tab:app-exp-map}
    \resizebox{\linewidth}{!}{
    \begin{tabular}{llll}
    \toprule
    \textbf{RQs} & \textbf{Method} & \textbf{Intent Dimensions} & \textbf{Setting} \\
    \midrule
    \textbf{RQ1} & FITD, MRCJ & DEA / SRA & Organization vs.\ accumulation \\
    \textbf{RQ2} & ActorAttack & DEA & Amplification components \\
    \textbf{RQ3} & MRCJ & SRA & Escalation shape \\
    \textbf{RQ4} & X-Teaming & SRA & Sub-paradigm progression \\
    \bottomrule
    \end{tabular}
    }
\end{table}

\subsection{Attack Models}
\label{app:attack_models}

Several attack implementations rely on an auxiliary LLM to perform attacker-side operations. 
Across our experiments, FITD~\cite{weng2025foot} uses such a model for rephrasing and bridge turns, ActorAttack~\cite{ren2024derail} for actor-specific pathways and self-talk, and X-Teaming~\cite{rahman2025x} for active planning, attack generation, and optimization. 
To control for differences in the underlying attacker-side model, we use \DeepSeek{} as the auxiliary attacker model in all experiments.

\subsection{Target Models}
\label{app:models}

\autoref{tab:app-models} lists the target models selected for each mechanism; MRCJ uses a broader set of open-source aligned models than the other experiments.

\begin{table}[t]
    \centering
    \caption{The summary of target models used in each experiment.}
    \label{tab:app-models}
    \resizebox{\linewidth}{!}{
    \begin{tabular}{ll}
    \toprule
    Experiment & Target models \\
    \midrule
    FITD &
    \LlamaTHREE, \DeepSeek, \GPTFourO, \LlamaTHREETHREE \\
    ActorAttack &
    \DeepSeek, \GPTFourO, \LlamaTHREETHREE \\
    X-Teaming &
    \DeepSeek, \GPTFourO, \LlamaTHREETHREE, \GeminiThree \\
    MRCJ &
    \VicunaSeven, \VicunaThirteen, \MISTRAL, \ChatGLM, \QWENTwo, \LlamaTHREE, \GemmaNine \\ 
    \bottomrule
    \end{tabular}
    }
\end{table}

\subsection{Datasets and Query Sets}
\label{app:datasets}

\autoref{tab:app-datasets} summarizes the released query sets used directly so that ablations vary interaction structure rather than harmful objectives.

\begin{table}[t]
\centering
\caption{Datasets and query sets used in the mechanism analyses.}
\label{tab:app-datasets}
\resizebox{\linewidth}{!}{
\begin{tabular}{lll}
\toprule
Dataset & Size / role & Used for \\
\midrule
JailbreakBench~\cite{chao2024jailbreakbench}
& 100 harmful behaviors
& FITD, X-Teaming \\
MUCD~\cite{MRCJ}
& Questions with malice levels 
& MRCJ auxiliary turns \\
AdvBench~\cite{zou2023universal}
& 50 final target questions
& MRCJ final targets \\
HarmBench~\cite{mazeika2024harmbench}
& 400 harmful queries
& ActorAttack \\
\bottomrule
\end{tabular}
}
\end{table}

\subsection{Evaluation Protocol}
\label{app:eval}

\noindent\textbf{Judge.} \GPTFourO{} serves as the automated judge across all experiments using the default JailbreakBench evaluation prompt~\cite{chao2024jailbreakbench}.
For each model response, it evaluates whether the generated content constitutes a successful jailbreak under the standardized JailbreakBench protocol.
The same judge prompt, model configuration, and evaluation procedure are used across all experiments to ensure comparability.

\noindent\textbf{Scoring and ASR.}
Responses are scored on a 1-to-5 harmfulness scale; a score $\geq 4$ counts as a successful jailbreak.
ASR is the fraction of target queries for which at least one response meets this threshold within the turn budget.
Where reported, the harmfulness score is the mean judge score across all target queries for a given condition.
For multi-candidate settings, a target is counted as successful if any candidate response reaches the threshold.

\noindent\textbf{Decoding.}
All target, attacker, and judge models use default decoding settings throughout unless otherwise specified by the released implementation.

\subsection{Per-RQ Condition Parameters}
\label{app:params}

\noindent\textbf{RQ\,1: Organization vs.\ Accumulation (Subsection~\ref{sec:org_vs_accum}).}
For the FITD-side experiment, we use the open-source implementation of Weng et al.~\cite{weng2025foot} with its default configuration.
All FITD-side conditions use a turn budget of 10.
In the Persistence condition, the attacker retries at most 10 times, with LLM-based rephrasing applied after refusals.
For the MRCJ-side experiment, auxiliary questions and malice-level labels follow the MRCJ release~\cite{MRCJ} without modification.
The final harmful targets are 50 questions from AdvBench~\cite{zou2023universal}.

\noindent\textbf{RQ\,2: Amplification Components (Subsection~\ref{sec:amplification}).}
RQ\,2 uses the open-source ActorAttack~\cite{ren2024derail} implementation with its default configuration to isolate the contribution of its key amplification components.
In the actor-count experiment, the number of actors varies from 1 to 5 while self-talk remains enabled, measuring the effect of increasing semantic diversity under a fixed reasoning mechanism.
In the self-talk ablation, the actor count is fixed at 3, and only the self-talk component is removed, with all other settings unchanged to ensure a controlled comparison.

\noindent\textbf{RQ\,3: Escalation Shape (Subsection~\ref{sec:escalation_shape}).}
\label{app:rq3}
RQ\,3 shares the target models, datasets, and evaluation protocol of the MRCJ side of RQ\,1.
The experiment fixes the number of escalation steps to 4, while the endpoint malice level and auxiliary-question construction follow the MRCJ default setting~\cite{MRCJ}. We measure query times (QT) as the number of question and answer exchanges between a user and the model required by a schedule; one exchange counts as one QT.

\noindent\textbf{RQ\,4: Sub-Paradigm Progression (Subsection~\ref{sec:subparadigm_progression}).}
\label{app:rq4}
All three conditions use the X-Teaming backbone~\cite{rahman2025x}, with \DeepSeek{} as the attacker LLM.
The turn budget is fixed to 7 across A1, A2, and A3, and the retry count is kept constant so that the comparison uses the same interaction budget.
The three conditions form a progressive configuration chain, as shown in \autoref{tab:xt-components}.
A1 executes a single fixed plan linearly (Fixed-strategy); A2 adds a candidate plan pool with an $\varepsilon$-greedy bandit selecting on scalar verifier scores (Adaptive-optimization); A3 replaces the bandit with an attacker LLM that reasons over the full interaction history (Agent-based).

\begin{table}[t]
    \centering
    \caption{Configuration matrix for the three main-chain conditions. Each column instantiates one SRA sub-paradigm; moving across columns jointly changes the planning resources and decision mechanisms characteristic of that sub-paradigm.}
    \label{tab:xt-components}
    \footnotesize
    \setlength{\tabcolsep}{4pt}
    \renewcommand{\arraystretch}{1.05}
    \begin{tabular}{@{}p{0.34\linewidth}p{0.18\linewidth}p{0.18\linewidth}p{0.20\linewidth}@{}}
        \toprule
        \textbf{Configuration aspect} & \textbf{A1} & \textbf{A2} & \textbf{A3} \\
        \midrule
        \textbf{Candidate plan pool}  & single plan & plan pool  & plan pool \\
        \textbf{Selection mechanism}  & none        & bandit     & LLM \\
        \textbf{Attacker model class} & none        & optimizer  & LLM agent \\
        \midrule
        \textbf{Sub-paradigm}         & Fixed       & Adaptive   & Agent \\
        \bottomrule
    \end{tabular}
\end{table}

\section{Method-Level Taxonomy of Multi-Turn Jailbreaks}
\label{app:method-level-taxonomy}

\noindent\textbf{\Circled{1} Method-level taxonomy exposes operational differences within and across intent dimensions.}
\autoref{tab:method_level_taxonomy} records each method's mechanism, interaction and observation scope, auxiliary capabilities, and code availability, revealing substantial differences among methods sharing a dimension.
Accordingly, a shared intent label denotes a common attack objective rather than an identical operational workflow. The table also makes explicit where a method relies on auxiliary models or released code, two factors that affect both reproducibility and the interpretation of defensive evaluations.

\noindent\textbf{\Circled{2} Single-trajectory interaction structures remain common, while harder detection regimes remain under-explored.}
Existing work concentrates on single trajectories with session-level observation, frequently combining DEA and SRA. TBA and SSA remain sparse despite requiring cross-session analysis. In particular, TBA requires cross-branch linkage, and both settings require defenses that aggregate evidence across longer interaction histories.
This concentration means that findings from the dominant settings may not transfer directly to attacks that link signals across sessions or branches. Evaluations in these underrepresented regimes should state the linkage mechanism and its observation cost explicitly, so that their results can be compared with those from localized interaction settings.

\end{document}